\documentclass[12pt,letterpaper]{article}
\pdfoutput=1
\usepackage{jheppub}
\usepackage{verbatim}

\usepackage{graphicx}

\usepackage{subfigure}
\input{epsf}
\usepackage{amsthm}

\def\bea{\begin{eqnarray}}
\def\eea{\end{eqnarray}}
\def\nn{\nonumber \\}
\def\rmd{{\rm d}}

\title{Scanning Tunneling Macroscopy, \\
Black Holes and AdS/CFT Bulk Locality}
\author[a]{Soo-Jong Rey}
\author{\hskip0.3cm \& \hskip0.3cm}
\author[b]{Vladimir Rosenhaus}
\affiliation[a]{School of Physics and Astronomy \& Center for Theoretical Physics \\
Seoul National University, Seoul 151-747 \rm KOREA}
\affiliation[b]{Department of Physics \& Berkeley Center for Theoretical Physics \\
University of California, Berkeley CA \rm USA}
\affiliation{\tt sjrey@snu.ac.kr \  vladr@berkeley.edu}

\abstract{ABSTRACT  }
\begin{document}

\abstract{
We establish resolution bounds on reconstructing a bulk field from boundary data on a timelike hypersurface.
If the bulk only supports propagating modes, reconstruction is complete. If the bulk also supports evanescent modes, local reconstruction is not achievable unless one has exponential precision in knowledge of the boundary data. Without exponential precision, for a Minkowski bulk, one can reconstruct a spatially coarse-grained bulk field, but only out to a depth set by the coarse-graining scale.
For an asymptotically AdS bulk, reconstruction is limited to a spatial coarse-graining proper distance set by the AdS scale.
AdS black holes admit evanescent modes.
We study the resolution bound in the large AdS black hole background and provide a dual CFT interpretation.
Our results demonstrate that, if there is a black hole of any size in the bulk, then sub-AdS bulk locality is no longer well-encoded in
boundary data in terms of  local CFT operators. Specifically, in order to probe the bulk on sub-AdS scales using only boundary data in terms of local operators, one must either have such data to exponential precision or make further assumptions about the bulk state.
}

\maketitle
\section{Introduction}
This paper concerns reconstructing the state of the anti-de Sitter (AdS) bulk from the conformal field theory (CFT) boundary. Finding which CFT quantities encode the bulk and, if so, in what ways have been actively pursued recently, e.g. \cite{Kab13,Heem12,Rose13,Kelly13, Lash13, Faulk13,Eng13, Hubeny13, Bal13, BouFre12,Leich13, Keeler13}. One standard approach is to start with local CFT data and use bulk equations of motion to evolve radially inward. The questions we seek to answer are: under what circumstances is this evolution possible, how deep into the bulk can we evolve, on what scales can the bulk be reconstructed, and what assumptions about the bulk state must be made?

We first address these questions in a simpler context: the electromagnetic field in Minkowski spacetime.
 Given boundary field data on a \textit{timelike} codimension-1 hypersurface (which we conveniently place at $z=0$), can the electromagnetic field be determined everywhere $(z>0)$? Suppose the bulk is filled with homogeneous air. The only solutions to the wave equation consistent with translation symmetry are propagating waves and  reconstruction is trivially achieved. On the other hand, suppose that the bulk is filled with air for $0 \le z<z_g$ and with glass for $z>z_g$. The translation symmetry being broken, there is now a new class of solutions: waves which are propagating for $z>z_g$ but evanescent for $0 \le z<z_g$. Evanescent modes are solutions with imaginary momentum. While legitimate solutions of the wave equation, these modes are forbidden in vacuo because of exponential growth at large $z$ and hence of non-normalizability. The presence of glass at finite $z$ cuts off this unboundedness and renders the mode permissible. Reconstructing the field inside the glass from measurements at $z=0$ is hopeless; a small mistake will get exponentially amplified. It would be like trying to measure the electromagnetic field inside a waveguide while standing a kilometer away. In fact, even  reconstruction of the field for $0 \le z<z_g$ from the boundary data at $z=0$ is no longer straightforward.
If we do not measure the evanescent modes (in the absence of assumptions on the form of the solution), we can not reconstruct the field anywhere. We can, however, measure the evanescent modes at $z=0$ to some extent without resorting to exponential precision. If we do, then the field, coarse-grained in $x$ over a scale $\sigma$, can be reconstructed, but only for $z<\sigma$.

We next address the reconstruction question in the context of the AdS/CFT correspondence: we consider a free scalar field in a fixed asymptotically AdS background. A background which is pure AdS, or a perturbation thereof, is like the electromagnetic field in the vacuum - exact reconstruction using boundary data is possible. Thus, local CFT operators give a probe of the bulk on the shortest of scales. A background with a small AdS black hole is like putting in a region with glass. The geometry in the black hole atmosphere, the region from $2M$ to $3M$, changes the boundary conditions and permits evanescent modes. Analogously to the case with glass, reconstructing the field in the atmosphere is hopeless. Reconstructing the field far from the black hole is possible to some extent and, like in the case of the electromagnetic field, requires measuring the evanescent data. The condition $z<\sigma$ translates into the ability to resolve the bulk on AdS scales and no shorter. The measurement of evanescent modes is the basis of the functioning of a Scanning Tunneling Microscope (STM). In this sense, the CFT is acting like an STM for the bulk at macroscopic scales.

It may seem puzzling that a small black hole deep in the bulk should have any impact on our ability to reconstruct the bulk close to the boundary. One may pretend the evanescent modes do not exist and work only with the propagating modes. This might be a good approximation for some states and for regions close to the boundary, but it is one that is violated by legitimate finite energy solutions having a significant amplitude for an evanescent mode.
Even in the Hartle-Hawking vacuum, the CFT Green's function $G_2(\omega, {\bf k})$ ( given by the Fourier transform of the finite temperature two-point correlator $\langle T \vert O(t, {\bf x}) O(0, {\bf 0}) \vert T \rangle$) is nonzero but exponentially small $\sim \exp (- \tilde{\alpha} |{\bf k}|/T)$ in the evanescent regime $k \gg \omega$. The evanescent modes are part of the spectrum of micro-canonical states, so even if one might opt to ignore them for two-point correlators, they necessarily contribute to any finite-temperature $n-$point function as intermediate states.

We have organized the paper as follows. In Section \ref{sec:1}, we formulate in Minkowski spacetime the problem of spacelike reconstruction from timelike boundary data. We show that reconstruction is exact in situations with only propagating modes but requires exponential precision in knowledge of boundary data in situations with evanescent modes. In the latter situation, reconstruction without exponential precision is possible but only at the cost of coarse-graining over directions parallel to the boundary and only out to a distance set by this averaging scale. In Section \ref{sec:AdS}, we formulate the reconstruction problem in AdS space and demonstrate that the situation is exactly parallel to the Minkowski spacetime counterpart. In a pure AdS background, all modes are propagating and the reconstruction is exact. In an AdS black hole background, evanescent modes open up near the black hole horizon and reconstruction requires exponential precision. Here again, reconstruction without exponential precision is possible but only at the cost of an AdS scale averaging over directions parallel to the boundary. In Section \ref{sec:CFT}, via the AdS/CFT correspondence, we discuss the impact of evanescent modes on bulk reconstruction from the CFT viewpoint. We show that in general precise determination of Green's functions at finite temperature requires exponential precision. In Appendix \ref{sec:EvaOptics} we review evanescent modes in optics, the principles of a microscope, and scanning tunneling (optical) microscopy (STM).

\section{Bulk reconstruction in Minkowski space} \label{sec:1}
\begin{figure}[tbp]
\centering
%\subfigure[]{
	\includegraphics[width=2in]{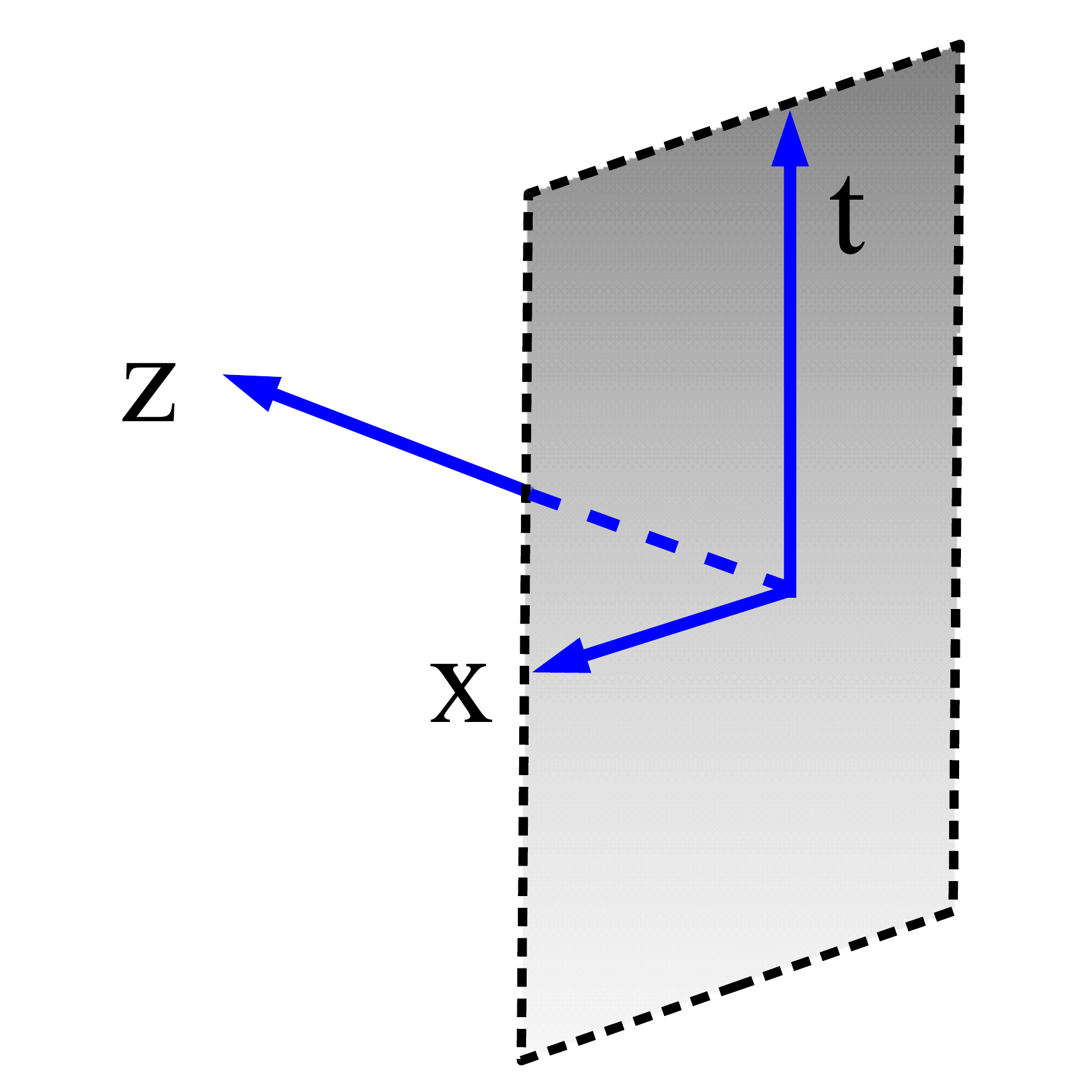}
%	}
\caption{\sl The setup of bulk reconstruction in $(2+1)$-dimensional Minkowski half space $\mathbb{R}^{2,1}_+$. We have a bulk field $\phi(x,t,z)$ obeying the wave equation, which we wish to reconstruct from the data $\phi(x, t, 0)$ on a timelike hypersurface  at $z=0$.} \label{fig:WavePlane}
\end{figure}

In this section, we pose the question: for a field $\phi(x,t,z)$ satisfying the wave equation in $(2+1)$-dimensional Minkowski spacetime \footnote{Extension to higher dimensions is straightforward and does not reveal any new physics.}, is the boundary data $\phi(x,t,0)$ specified at a timelike hypersurface $z=0$ sufficient to reconstruct the field $\phi(x,t,z)$ everywhere?  (See the setup shown in Fig. \ref{fig:WavePlane}.) The wave equation admits solutions with both real and complex momentum. The solutions with real momentum are the propagating waves, $e^{\pm i k_z z}$. If the space is everywhere homogeneous, such as pure Minkowski space, then these are the only admissible delta-function normalizable modes. In this case, reconstruction of $\phi(x,t,z)$ from the boundary data $\phi(x',t',0)$ works perfectly --- we show in Sec.~\ref{sec:prop} that the smearing function $K(x,t,z|x',t')$, whose convolution with $\phi(x',t',0)$ yields $\phi(x,t,z)$, is well-defined everywhere in the bulk. On the other hand, if the space is inhomogeneous by, for instance, having a spatially varying index of refraction, then modes with imaginary momentum can become permissible. Instead of propagating, these modes grow exponentially in the $z$ direction: $e^{ \pm \kappa_z z}$. They are known as ``evanescent modes''. Our goal is to point out that the evanescent modes cause serious difficulties in reconstructing the field anywhere from given boundary data.

\subsection{Success with propagating modes} \label{sec:prop}
We consider the wave equation in $(2+1)$-dimensional Minkwoski spacetime $\mathbb{R}^{2,1}$:
\begin{equation} \label{eq:WaveMink}
\left(-\partial_t^2 + \partial_x^2 + \partial_z^2\right)\ \phi(x,t, z) = 0 .
\end{equation}
We will be interested in reconstructing $\phi(x,t,z)$ from boundary data $\phi(x, t, 0)$ specified on a timelike hypersurface $b$ at $z=0$. To accomplish this, we decompose the solutions to (\ref{eq:WaveMink}) in terms of the propagating wave basis:
\begin{equation} \label{eq:phiModes}
\phi(x,t,z) = \int\int{\rmd k_x \ \rmd \omega \ \phi(k_x, \omega)\ e^{- i k_x x - i k_z z - i \omega t}},
\end{equation}
where the delta-function normalizability condition puts
\begin{equation}\label{eq:kz}
k_z \equiv \sqrt{\omega^2 - k_x^2} \in \mathbb{R}^+ \quad \rightarrow \quad |\omega| \ge |k_x|.
\end{equation}
We chose $k_z$ to be positive, as we will for simplicity assume there are only left-movers.\footnote{Assuming there are only left-movers allows us to connect more directly with the analogous problem in AdS space. The more general case in Minkowski space for which right-movers are also included is a straightforward extension and requires specifying $\partial_z \phi(x,t,z=0)$ in addition to $\phi(x,t,z=0)$ at the boundary $b$.} The decomposition (\ref{eq:phiModes}) is slightly nonstandard as $\omega$ is one of the independent variables as opposed to $k_z$. This is a convenient choice here as the boundary data is specified on a timelike hypersurface. The Fourier transform of $\phi(x,t,z=0)$ yields
\begin{equation} \label{eq:phiFou}
\phi(k_x,\omega) = \int\!\!\!\int_{b} {\rmd x' \rmd t'\ e^{i k_x x' + i \omega t'} \phi(x',t',0)}.
\end{equation}
Inserting (\ref{eq:phiFou}) into (\ref{eq:phiModes}),
\begin{equation} \label{eq:phiK}
\phi(x,t,z) = \int\!\!\!\int_b {\rmd x' \rmd t'\ K(x,t,z| x',t') \phi(x',t',0)}.
\end{equation}
Here, $K$ is the smearing function given by
\begin{eqnarray} \label{eq:K}
K(x,t,z|x',t') &=& \int_{-\infty}^\infty \rmd \omega \int_{\vert k_x \vert \le \vert \omega \vert} \rmd k_x  \ { e^{- i\omega (t-t')}  e^{-i k_x (x-x')} e^{-i \sqrt{\omega^2 - k_x^2}\ z}} \nn
&=&\int_{-\infty}^\infty \rmd k_x \int_{\vert \omega \vert \ge \vert k_x\vert} \rmd \omega  \ { e^{- i\omega (t-t')}  e^{-i k_x (x-x')} e^{-i \sqrt{\omega^2 - k_x^2}\ z}}
.
%K(x,t,z|x',t') = \int{d\omega \  K_{\omega}(x,z|x') e^{-i \omega(t-t')}}.
\end{eqnarray}
The integral in (\ref{eq:K}) is convergent, so (\ref{eq:phiK}) realizes our goal of reconstructing $\phi(x,t,z)$ in terms of the boundary data $\phi(x,t,0)$. This is not surprising, as for every mode the field at some value of $z$ is related to the field at $z=0$ by the phase-factor $e^{i k_z z}$ associated with the translation, where $k_z$ is given by (\ref{eq:kz}).

It will be useful for later to first reconstruct per each monochrome $\omega$ mode and then combine them together. We then use
\begin{equation}
\phi_{\omega}(x,0) \equiv \int_{-\infty}^\infty { \rmd t\ e^{i \omega t}\ \phi(x,t,z=0)}
\end{equation}
to reconstruct $\phi_{\omega}(x,z)$,
\begin{equation}
\phi_{\omega}(x,z) = \int_{-\infty}^\infty { \rmd x'\ K_{\omega}(x,z| x')\ \phi_{\omega}(x',0)}
\end{equation}
where
\begin{equation} \label{eq:Komega}
K_{\omega}(x,z|x') = \int_{\vert k_x \vert \le \vert \omega\vert} {\rmd k_x \ e^{-i k_x (x-x')} e^{-i \sqrt{\omega^2 - k_x^2}\ z}}.
\end{equation}
The full smearing function is then recovered through the spectral sum
\begin{equation} \label{eq:KKomega}
K(x,t,z|x',t') = \int_{-\infty}^\infty {\rmd\omega \  K_{\omega}(x,z|x') e^{-i \omega(t-t')}}.
\end{equation}

\subsection{Reality with evanescent modes}
In the above discussion, we restricted $k_z$ to be real-valued as a consequence of the mode's delta-function normalizability condition. If one were to consider $k_x> \omega$ such that $k_z$ is imaginary: $k_z \equiv i \kappa_z$, a mode would take the form
\begin{equation} \label{eq:Evan}
e^{- i k_x x} e^{- i \omega t} e^{\kappa_z z} \qquad (z \ge 0)~.
\end{equation}
This solution diverges at large $z$, and is therefore not permissible. We were thus correct to discard it. However, if the Minkowski space is inhomogeneous in the $z$-direction, often solutions such as (\ref{eq:Evan}) become permissible. In Appendix \ref{sec:EvaOptics}, we give two situations which generate evanescent modes: (1) a wave traveling in a medium with a $z$-dependent index of refraction, and (2) a wave scattering off a material  that has an $x$-dependent transmission coefficient and is located at some fixed $z$.

If evanescent modes are present, then they pose a serious challenge for reconstructing the field from the $z=0$ boundary data. Unlike propagating modes, evanescent modes will have an exponentially suppressed imprint on the $z=0$ boundary compared to their value at $z > 0$. This exponential behavior renders the reconstruction procedure of Sec.~\ref{sec:prop} inapplicable: the smearing function $K$ (\ref{eq:K}) is ill-defined because of the divergence from the $k_x\gg \omega$ region of integration. For the rest of this section, we simply assume that evanescent modes are present and do not inquire as to their origin. An exemplary situation helpful to keep in mind is the one where the index of refraction of the background changes at some large value of $z>0$. For simplicity, we will assume the change made at large $z$ is such that all possible evanescent modes are produced so that the mode solutions now contain all values of $k_x$ regardless of $\omega$. We would like to understand the impact this has on reconstructing the field from boundary data.

We can take two approaches in dealing with evanescent modes. We may ignore the modes from the outset, contending that they are not propagating. Or we may include the modes. We now argue that in both cases we will face unavoidable limitations on the resolving power of the reconstruction.

\subsection{Ignoring evanescent modes} \label{sec:eva}
Consider first the approach of ignoring the evanescent mode data at the boundary: we will not try to extract their coefficient from the boundary data provided at $z=0$. In the procedure of Sec.~\ref{sec:prop}, we are now in the situation that the Fourier decomposition of $\phi(x, t, z)$ (\ref{eq:phiModes}) extends over all $k_x$ and $\omega$, but we truncate the Fourier decomposition of the smearing function $K$ in (\ref{eq:K}) to $|\omega| \ge |k_x|$ .
Reconstruction of $\phi(x,t,z)$ can be done monochromatically per each frequency $\omega$, first reconstructing $\phi_{\omega}(x,z)$ and then combining them to get $\phi(x, t, z)$. Reconstruction of $\phi_{\omega}(x,z)$ is the same problem as the one encountered in optics when one tries to resolve features of a sample by shining monochromatic light on it. (For those who need to refresh optics, consult Appendix \ref{sec:EvaOptics}.) Ignoring the evanescent modes is the standard assumption in optical microscopy: the detector (playing the role of the boundary) is far from the sample, so the magnitude of the evanescent modes at the screen is exponentially small and is zero for all practical purposes. For this reason, we only have knowledge of features of the sample, $T(k_x)$, for $|k_x|\le |\omega|$. This is the standard result we would expect: detecting light of frequency $\omega$, we can probe the features of a sample but only on scales larger than the resolution power set by $\omega^{-1}$.

Specifically, we assume the boundary data specified at the $z=0$ hypersurface tells us nothing about the coefficients $\phi_{\omega} (k_x)$ for $|k_x| \ge |\omega|$. Clearly, we can then not hope to confidently reconstruct $\phi_{\omega}(x,z)$, as the coefficient $\phi_{\omega} (k_x)$ for the large $k_x$ could be arbitrarily large. To ameliorate this, we may ask for reconstructing the field $\phi_{\omega}(x,z)$ coarse-grained over $x$ with the Gaussian window function of resolution scale $\sigma$. We denote this coarse-grained field as $\phi_{\omega}^{\sigma}(x,z)$:
\begin{equation} \label{eq:phiSigma}
\phi_{\omega}^{\sigma}(x,z) = \int_{-\infty}^\infty {\rmd x' \ e^{- (x-x')^2/\sigma^2}} \ \phi_{\omega}(x',z) \ .
\end{equation}
We would expect reconstructing $\phi_{\omega}^{\sigma}(x,z)$ should only require knowledge of modes with $|k_x| \lesssim |\sigma|^{-1}$. To verify this, we rewrite (\ref{eq:phiSigma}) as
\begin{equation}
\phi_{\omega}^{\sigma}(x,z) = \int{\rmd k_x\ \phi_{\omega}(k_x,z)\ e^{- k_x^2 \sigma^2}\ e^{-i k_x x}} .
\end{equation}
Unless $\phi_{\omega}(k_x, z)$ grows exponentially in $k_x^2$ for large $k_x$, its value is irrelevant for $|k_x|\gtrsim |\sigma|^{-1}$. Therefore, if we are only interested in reconstructing the field $\phi_{\omega}(x,z)$ coarse-grained over $x$ with resolution scale $\sigma$, then we can reconstruct it using only the propagating mode data at the $z=0$ boundary {\sl as long as} $|\omega| \gtrsim 1/\sigma$ (and provided a reasonable assumption is made about the behavior of the $|k_x|\gg |\omega|$ modes).

All seems well, but there is a problem. To reconstruct $\phi^{\sigma}(x,t,z)$ for {\sl all} time $t$, we need $\phi_{\omega}^{\sigma}(x,z)$ for all $\omega$. Yet, there is no $\sigma$ for which this condition will be satisfied: for any $\sigma$, there is an interval of missing $\omega$, $|\omega|<1/\sigma$. This means that, with the assumption we made that we ignore the evanescent mode data at $z=0$, we can not possibly reconstruct the temporal evolution of the field at any $z$ location, regardless of how large a coarse-graining resolution $\sigma$ in $x$ we are willing to compromise for. In short, for reconstruction relying only on propagating mode data, coarse-graining over $x$ achieved reconstructability over $z$ but sacrificed reconstructability over $t$.
%-for microscope, ok. since at fixed omega. for us, disaster since need all omega.

\subsection{Dealing with evanescent modes} \label{sec:capture}
%In this section we will not make the assumption of Sec.~\ref{sec:eva} that data at $z=0$ tells us nothing about the coefficient of any of the evanescent modes assumption and
Consider next the approach of retaining all the evanescent mode data at the boundary $b$.
If we wish to resolve the field on a scale $|\Delta x| \simeq \sigma$, we expect to require knowledge of the field profile
with $k_x \lesssim \sigma^{-1}$. But then, evanescent modes with $\omega=0$ and $k_x = \sigma^{-1}$ behave like $e^{z/\sigma}$, and we would expect we can determine $\phi^{\sigma}(x,t,z)$ only for $z\lesssim \sigma$. At bulk locations $z$ larger than $\sigma$, the evanescent modes grow exponentially large compared to what we have access to on the boundary. We will show below that one would require exponential accuracy in the knowledge of the $z=0$ boundary data to reconstruct the field at bulk regions as deep as $z \gg \sigma$. To foreshadow considerations in Sec.~\ref{sec:AdS} for AdS space, let us mention that the criterion $z\lesssim \sigma$ will turn into $\sigma_{\text{proper}}(z) \gtrsim L_{AdS}[1 + \varepsilon(z)]$, where $\sigma_{\text{proper}(z)}$ is the proper distance at a bulk location $z$ corresponding to the coordinate interval $\sigma$, up to a correction factor $\varepsilon(z)$ that depends on details of the bulk.

We now redo the computation of section~\ref{sec:prop} but with evanescent modes taken into account. The monochrome
field $\phi_{\omega}(x,z)$ is expressed in terms of $\phi_{\omega}(x,0)$ through $K_{\omega}$ of (\ref{eq:Komega}). In the expression (\ref{eq:Komega}) for $K_{\omega}$, we must integrate over all $|k_x|< \infty$. We split this integral into contributions of the propagating modes and of the evanescent modes, respectively,
\begin{equation} \label{eq:Kfull}
K_{\omega}(x,z| x') = \int_{-\omega}^{\omega}{\rmd k_x\ e^{- i k_x (x-x')} e^{- i \sqrt{\omega^2 - k_x^2}\ z}} + \int_{|k_x|>\omega}{\rmd k_x\ e^{- i k_x (x-x')}\ e^{\sqrt{k_x^2 - \omega^2}\ z}}.
\end{equation}
The second integral is badly divergent. Thus, the evanescent modes have obstructed our ability to reconstruct the field in the bulk. As before, we instead ask for reconstructing the more physical quantity: the field coarse-grained over $x$ with a Gaussian window function. We have
\begin{equation}
\phi_{\omega}^{\sigma}(x,z) = \int_b {\rmd x'\ K_{\omega}^{\sigma}(x,z| x')\ \phi_{\omega}(x',0)} \,
\end{equation}
where
\begin{equation} \label{eq:Ksigma}
K_{\omega}^{\sigma} (x,z| x') = \int_b {\rmd \bar{x}\ e^{-(x-\bar{x})^2/\sigma^2} K_{\omega}(\bar{x},z| x')} \ .
\end{equation}
Inserting (\ref{eq:Komega}) into (\ref{eq:Ksigma}) yields
\begin{equation} \label{eq:Kreg}
K_{\omega}^{\sigma} (x,z| x') = \int{\rmd k_x\ e^{- i k_x (x-x')}\ e^{-i \sqrt{\omega^2 - k_x^2}\ z}\ e^{- k_x^2 \sigma^2}} \ .
\end{equation}
Asking for the field smeared over $\sigma$ in the $x$ direction thus amounts to coarse-graining the smearing function (\ref{eq:Kfull}) over momentum $k_x$ with the Gaussian window function $\exp(-k_x^2 \sigma^2)$ to suppress large $k_x$.
% Maybe evaluate (\ref{eq:Kreg}) exactly??
The $k_x\gg \omega$ part of the integral in (\ref{eq:Kreg}) gives
\begin{equation} \label{eq:Exp}
[e^{ - 2 i (x-x') z/\sigma^2} \cdot e^{z^2/\sigma^2}] \ e^{-(x-x')^2/\sigma^2} \ .
\end{equation}
Having smeared out in the $x$-direction, let's examine the behavior in the bulk $z$-direction. As measured at deeper bulk regions $z \gg \sigma$, this function oscillates rapidly in phase {\sl and} grows exponentially in amplitude. So, to determine the $\sigma-$grained field at a $z$ greater than $\sigma$, one needs to measure the boundary data $\phi_{\omega}(x,z=0)$ with exponential precision. In short, for reconstruction retaining evanescent mode data, coarse-graining over $x$ ameliorated reconstructability over $z$ for a depth of order the coarse-graining scale or so.

\subsection{Window function}
In both approaches to dealing with evanescent modes, bulk reconstruction required a choice of a window function for the $x$-direction to regularize the divergence in (\ref{eq:Kfull}).
%Another difficulty is the sensitivity to the precise way in which we choose to coarse-grain in the $x-$direction.
In (\ref{eq:Kreg}), we chose a Gaussian window function to achieve the regularization. What about other choices?
One might try a hard-wall window function, $\Theta(\sigma^{-1}-|k_x|)$, which gives perfect regularization, and whose $x$-space window function takes the form
\begin{equation}\label{eq:sinFil}
\frac{\sin\left(\frac{x-x'}{\sigma}\right)}{x-x'} ~.
\end{equation}
 One might also try a Laplace window function, $e^{- k_x \sigma}$, but this function would be insufficient to regulate the divergence. Details of reconstruction certainly depend on the choice of the window function, but the fact that we are limited by resolution bounds does not depend on the choice.

The choice of the window function is also a practical matter for the Scanning Tunneling Microscopy (STM). The basis of STM is
the measurement of evanescent modes. In the problem of resolving the features of a sample (Appendix \ref{sec:EvaOptics}),
the location of the sample is held fixed and bringing the STM probe needle a distance $\sigma$ close to the sample allows
image resolution on a scale $\sigma$. From the optics perspective, determining the spatial features of a sample regardless
of the frequency with which it is illuminated, as an STM allows, is an enormous achievement. We have shown that the ability
to do this in the STM context translates into our ability to reconstruct the bulk field $\phi^{\sigma}(x,t,z)$ for $z<\sigma$  from boundary data. While the precise depth to which reconstruction is possible would certainly depend on the chosen window function, the fact that the STM probes to a depth set by the image resolution scale is independent of the choice.

One should note that the coarse-grained smearing function, for any coarse graining other than the hard wall choice (\ref{eq:sinFil}),
makes little distinction between $\sigma$ less than or greater than $1/\omega$. This appears in tension with the standard
assumption (reviewed in section~\ref{sec:eva}) that, for $\sigma> 1/\omega$, reconstruction should still be possible even while
ignoring the evanescent modes. However, regardless of $\omega$, $K_{\omega}^{\sigma}$ has the same behavior coming from
$k_x\gg \omega$ and leading to the same complications with reconstruction for $z\gtrsim \sigma$.
The resolution is that the smearing function makes no assumption regarding the high $k_x$ behavior of the field we aim
to reconstruct, whereas our previous argument that modes with $k_x>\sigma$ are not needed for the $\sigma$-grained field
relied on a (very reasonable) assumption about the high $k_x$ behavior of the field.

\subsection{Conclusions}
Let's summarize what we have learned so far. Our question of interest has been whether we can reconstruct the field $\phi(x,t,z)$ using the data $\phi(x,t,z=0)$ at the $z=0$ timelike hypersurface. We assumed we can do this reconstruction monochromatically, reconstructing $\phi_{\omega}(x,z)$ from $\phi_{\omega}(x,0)$. If the medium is homogeneous and only propagating modes are present, then the reconstruction works flawlessly and is given by ({\ref{eq:phiK}, \ref{eq:K}}). If, however, the medium is inhomogeneous and evanescent modes are present, then reconstructing $\phi(x,t,z)$ point by point in the bulk is not achievable. We could instead reconstruct $\phi_{\omega}(x,z)$ coarse-grained in $x$ with a $\sigma$-sized resolution, $\phi_{\omega}^{\sigma}(x,z)$. This reconstruction can be done while not measuring any of the evanescent modes, but only for $\omega \gtrsim \sigma^{-1}$. Since reconstructing $\phi^{\sigma}(x,t,z)$ for all time $t$ requires reconstructing $\phi_{\omega}^{\sigma}(x,z)$ for all $\omega$, we are unable to reconstruct $\phi^{\sigma}(x,t,z)$ for any $z$.
If, on the other hand, we do include the evanescent modes, \textit{the reconstruction of $\phi^{\sigma}(x,t,z)$ can be done but only for  the bulk depth } $z\lesssim \sigma$. This is because reconstructing for $z\gtrsim\sigma$ would require exponential accuracy in measuring the value of $\phi(x,t,0)$.

\section{Bulk reconstruction in AdS space} \label{sec:AdS}
We now turn to AdS space and repeat the bulk reconstruction analysis. In short, we will find that the conclusion is exactly the same as in the flat Minkowski space case, except that we now need to understand the resolving power in proper distances.
In Sec.~\ref{sec:local},  we make some general remarks regarding the relation between restricting the boundary data
to local CFT operators and reconstructing a bulk AdS field  from this data. In Sec.~\ref{sec:nearBdy} we focus
on reconstruction of the near-boundary region of the bulk. In Sec.~\ref{sec:genBack},
we consider evolving deeper into the bulk and find how the resolving power is modified.

\subsection{Local CFT operators as boundary data} \label{sec:local}
The AdS/CFT dictionary in extrapolate form \cite{BanDoug98} relates the boundary limit of a bulk operator $\hat{\phi}$ to
a local CFT operator $O$:
\begin{equation} \label{eq:Ext}
\text{lim}_{z\rightarrow 0}\ z^{-\Delta}\ \hat{\phi}(x,t,z) = O(x,t).
\end{equation}
Hereafter, the only boundary CFT data we will consider is that of local CFT operators.
We also hold the CFT Hamiltonian fixed, corresponding to the restriction that all non-normalizable modes of the bulk field $\phi$ are
turned off. So, different states of the bulk field $\phi$ correspond to exciting different normalizable modes.
The CFT operator $O(x,t)$ dual to the bulk field $\phi$ has a scaling dimension $\Delta$ set by the mass of the bulk field $\phi$, $\Delta (\Delta -d) =m^2$. In non-vacuum states, the CFT operators acquire nonzero expectation values.

Defining the boundary tail through $\text{lim}_{z\rightarrow 0}{\phi}(x,t,z) = \phi_0(x,t)z^{\Delta}$, one would like to
express the field $\phi(x,t,z)$  in the bulk in terms of the boundary tail $\phi_0(x',t')$. This question is a nonstandard boundary-value problem of evolving
boundary hyperbolic data at a timelike hypersurface into the bulk along a spacelike `radial' direction. This is precisely the sort of problem we addressed in section \ref{sec:1} in the simpler context of flat Minkowski space.
In normal Cauchy evolution of an initial-value problem, we take a Fourier transform with respect to the spatial direction
$x,z$ of elliptic data on the $t=0$ Cauchy surface to obtain their spectral initial values. Here, we must work with hyperbolic data on the $z=0$ timelike hypersurface. So, as in section 2, instead of doing a Fourier transform with respect to $z$,
we do one with respect to boundary time $t$ and obtain spectral boundary values.

Suppose complete reconstructability of the bulk is possible. Then, the bulk operator $\hat{\phi}(x,t,z)$ at a bulk location $z$ is a linear combination of the CFT operator $O(x, t)$,
smeared over the boundary hypersurface $b$:
\begin{equation} \label{eq:KAdS}
\hat{\phi}(x,t,z) = \int_b {\rmd x'\ \rmd t' K(x,t,z|x',t')\ O(x',t')} + \cdots \ .
\end{equation}
Here, the ellipses refer to nonlinear interactions in the bulk. Although (\ref{eq:KAdS}) is an operator statement,
as a result of (\ref{eq:Ext}), finding $K$ is reduced to a classical field theory  problem.
In the limit of weakly interacting bulk dynamics (corresponding to arbitrarily large CFT central charge), the nonlinear bulk
interactions denoted by the ellipses in (\ref{eq:KAdS}) are negligible and the reconstruction can be done mode by mode.
In what follows, we will only be interested in the leading-order term of (\ref{eq:KAdS}). In the bulk, we then have a fixed background on which the non-interacting bulk field $\phi(x,t, z)$ evolves according to the AdS wave equation.
%The problem we face is: given $\phi_0(x,t)$ on timelike boundary hypersurface, can we obtain $\phi(x,t,z)$ through the bulk?
%This is precisely the same sort of question we addressed in Sec.~\ref{sec:1} in the simpler context of Minkowski space.

Indeed, we can construct a smearing function for pure AdS space, as was done
in \cite{HamKab05,HamKab06}. Using perturbation theory, we can also construct a smearing function for an asymptotically
AdS space perturbatively connected to pure AdS space (such as the one with a planet). In cases when a smearing function exists, through (\ref{eq:KAdS}),  the operators $O$ provide us with a probe of bulk locality on as short of scales as the semiclassical equations of motion are valid.
In cases when a well-defined smearing function $K$ can be constructed,  (\ref{eq:KAdS}) can be used to express bulk $n$-point correlators of $\hat{\phi}$ in terms of boundary CFT correlators of $O$.

We stress that having a well-defined smearing function is a more stringent requirement than
simply having an algorithm for determining the bulk field $\phi(x,t,z)$ from given conformal boundary data $\phi_0(x',t')$.
For instance, one might suppose that, for any particular bulk solution, even if $\phi_0(x',t')$ is extremely small,
one can pick an appropriate resolution for one's measuring device so as to see it and reconstruct $\phi(x,t,z)$.
However, it could be that no matter how good a resolution one picks, there always exist field configurations having
a near-boundary imprint $\phi_0(x',t')$ that is below the resolution power of one's measuring apparatus.
In such a case, one does not have a well-defined smearing function as its existence implies a state-independent way of reconstruction.
In a sense, one has to pick {\sl a priori} the resolution power without preconceived knowledge of which field configurations
will be under consideration.  Indeed, in some cases, there is no smearing function  \cite{Leich13}. A black hole background
is such a case -- there are modes with $\omega \ll l$ (where $l$ is the linear or angular momentum measured in units
of the AdS-scale) whose relative boundary imprint $e^{-l}$ is exponentially small.

A remark is in order regarding our working assumption on the boundary data. Nothing we have said so far regarding reconstruction has made use of there actually existing a dual CFT,
except that we assumed from the outset that the boundary conformal data is spanned by the set of local CFT operators.
We can consider a collection of near-boundary observers who reconstruct the bulk field, their limitation of being confined to
$z=0$ is overcome by making measurements over an extended period of time. This setup is the equivalent of the statement that a complete set of local CFT operators $O(x, t)$ smeared over
space and time is sufficient to reconstruct the bulk. It is important to note that this is a working assumption.
The CFT is equipped with many quantities that local near-boundary observers are not.
For instance, one could reconstruct the bulk by measuring CFT Wilson loops \cite{Rey:1998ik, Maldacena:1998im, PolSus99} or entanglement entropy \cite{RyuTak06}.
An important question is in what situations data provided by expectation values of local CFT operators $O(x, t)$ is
not sufficient to reconstruct the bulk and these other CFT quantities must be invoked.
In situations when local operators $O(x, t)$ are insufficient to exactly probe bulk locality,
we would like to understand if the $O(x, t)$'s are at least sufficient to probe the bulk fields coarse-grained over some scale.

\subsection{Bulk reconstruction near the boundary} \label{sec:nearBdy}
Our goal is to study  bulk reconstructability for general bulk spaces that asymptote to pure AdS space near the timelike boundary.
Let's first consider the case of reconstructing the bulk field at locations close to the boundary.
There, the metric is approximated by that of pure AdS space, so we should be able to make universal statements concerning
the reconstruction. The mode solutions on a general asymptotic AdS space can be classified into propagating and evanescent,
depending on whether the bulk radial momentum-like quantum number is real or imaginary.
If the bulk supports only propagating modes such as in pure AdS space, one can show that $\phi(x,t,z)$ can be
reconstructed exactly. On the other hand, if somewhere in the bulk there is a significant change in the geometry, then
evanescent modes may become permissible.
%Recall that, in the case of the wave equation in flat Minkowski space in Sec.~\ref{sec:1},
%a change in the index of refraction of the medium in the bulk caused evanescent modes to appear.

In this section, in situations where evanescent modes are present, we will show that
(1) if we completely ignore boundary data from the evanescent modes,
the bulk can not be reconstructed anywhere or on any scale,
and (2) if we include boundary data from the evanescent modes, but not to exponential precision,
\textit{we are able to reconstruct the bulk, but only on AdS-size scales and not shorter;
viz. we can reconstruct $\phi^{\sigma}(x,t,z)$ for $\sigma_{proper}>L_{AdS}$.}

%\subsubsection{Propogating Modes}
%\subsubsection{Evanecsent Modes}

\subsubsection*{AdS smearing function}
Let's first work out the explicit functional form of the smearing function.
Consider the massive scalar wave equation
\begin{equation} \label{eq:wave}
\frac{1}{\sqrt{g}}\partial_{\mu}(\sqrt{g}g^{\mu \nu} \partial_{\nu} \phi) - m^2 \phi =0,
\end{equation}
in the Poincar\'e patch of $(d+1)$-dimensional AdS space~\footnote{Hereafter, we shall suppress the AdS scale $L_{\rm AdS}$, and reinstate it in some of the final expressions.}  :
\begin{equation}
\rmd s^2 = \frac{-\rmd t^2+ \rmd {\bf x}^2 + \rmd z^2}{z^2} .
\end{equation}
The modes are
\begin{equation} \label{eq:AdSModes}
\phi_{\omega, {\bf k}_x} = 2^{\nu} \Gamma(\nu+1) q^{-\nu} \ z^{d/2}\ e^{- i {\bf k}_x \cdot {\bf x} - i\omega t}\ J_{\nu}(q z),
\end{equation}
where $q = \sqrt{\omega^2-{\bf k}_x^2}$ is like a bulk radial momentum and $\nu^2 = m^2 + (d/2)^2$. In (\ref{eq:AdSModes}),
we normalized the modes so that they approach plane waves of unit amplitude near the boundary:
\begin{equation}
z^{-\Delta} \phi_{\omega, {\bf k}_x} \rightarrow e^{- i {\bf k}_x \cdot {\bf x} - i\omega t} \ \ \ \text{as } \  z\rightarrow 0 ~.
\end{equation}

We would like to express the bulk field $\phi({\bf x},t,z)$ in terms of the conformal boundary data $\phi_0({\bf x}, t)$ at the $z=0$ hypersurface $b$,
\begin{equation} \label{eq:phiKAdS}
\phi({\bf x},t,z) = \int_b {\rmd {\bf x}' \rmd t'\ K({\bf x},t,z| {\bf x}',t') \ \phi_0({\bf x}',t')} ~.
\end{equation}
This equation is essentially identical to the one used in the context of reconstruction in Minkowski space (\ref{eq:phiK}).
The only difference is that since in AdS space all modes universally die off near the boundary,
on the right of (\ref{eq:phiKAdS}) we used the conformal data of the field
$\phi_0({\bf x}',t') \equiv \text{lim}_{z \rightarrow 0} z^{-\Delta} \phi({\bf x}',t',z)$.
We construct the smearing function $K$ in the same way as in section~\ref{sec:prop}: a Fourier transform
of the hyperbolic boundary $b$ allows us to extract from $\phi_0({\bf x},t)$ the Fourier coefficients of all modes.
This yields the smearing function $K({\bf x},t,z| {\bf x}',t')$ given by
\bea
K({\bf x}, t, z|{\bf x}',t') = \int_{-\infty}^\infty \rmd \omega \ K_\omega({\bf x}, z \vert {\bf x}')  e^{- i \omega (t - t')},
\eea
where
\begin{equation} \label{eq:KomegaAdS}
K_{\omega}({\bf x},z|{\bf x}') = 2^{\nu} \Gamma(\nu+1) \int_{-\infty}^\infty {\rmd {\bf k}_x\ e^{- i {\bf k}_x \cdot
({\bf x}-{\bf x}')} z^{d/2}\ q^{-\nu}  J_{\nu}(q z)}~.
\end{equation}
We reemphasize that (\ref{eq:KomegaAdS}) is the counterpart of (\ref{eq:Komega}), except that AdS modes have different behavior in the radial $z$-direction (\ref{eq:AdSModes}) as compared to Minkowski modes.
The smearing function (\ref{eq:KomegaAdS}) accounts for this difference.

\subsubsection*{Propagating modes}
\begin{figure}[tbp]
\centering
%\subfigure[]{
	\includegraphics[width=2.5in]{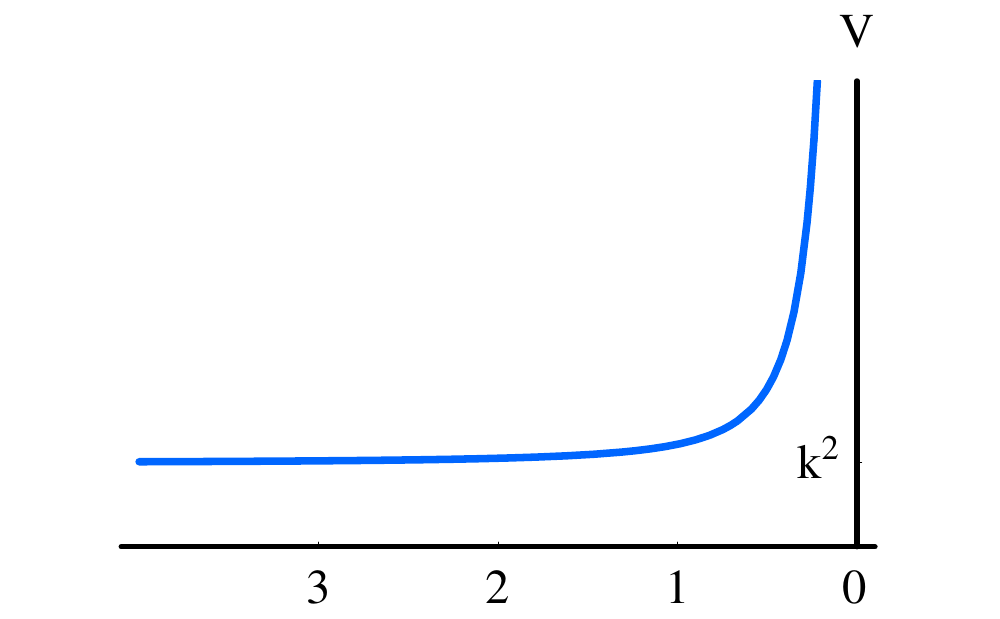}
%	}
\caption{\sl The bulk profile of the modes is found from solving the Schr\"odinger equation in the above effective potential $V(z)$. Modes with $\omega^2>{\bf k}_x^2$ are propagating, whereas those with $\omega^2<{\bf k}_x^2$ are evanescent. In pure AdS space, evanescent modes are forbidden due to their exponential divergence at large $z$. However, evanescent modes become permitted if there is a change in the effective potential in the large $z$ interior so as to cause the potential to dip below ${\bf k}_x^2$.} \label{fig:poinpatch}
\end{figure}
Let's analyze in what circumstances the smearing function is well-defined. We will find that, as in the Minkowski space analysis of section 2,
the AdS smearing function (\ref{eq:KomegaAdS}) is well-defined if there are only propagating modes.
From (\ref{eq:wave}), we see that the $z$-component
of the AdS modes, $\phi_{\omega {\bf k}_x} = u(z) z^{\frac{d-1}{2}} e^{-i {\bf k}_x \cdot {\bf x} - i \omega t}$,  satisfies
\begin{equation}
-\frac{d^2 u(z)}{d z^2} + V(z) u(z) = \omega^2 u(z),
\end{equation}
where
\begin{equation}
V(z) = {\bf k}_x^2 + \frac{\nu^2 - 1/4}{z^2} ~.
\end{equation}
The effective potential is plotted in Fig.~\ref{fig:poinpatch}. Toward the $z =0$ boundary, the potential is confining.
Toward $z \rightarrow \infty$,
the potential becomes flat and we recover the behavior of Minkowski modes. Indeed, in (\ref{eq:AdSModes}),
%We adopt the equations used in Sec.~\ref{sec:1} for the Minkowski problem to the current AdS case.
%We construct the smearing function $K_{\omega}$ (\ref{eq:Komega}) adopted to the AdS modes,
the Bessel function asymptotes for $q z \gg \nu$ to a plane wave:
\begin{equation} \label{eq:J}
J_{\nu}(q z) \sim \frac{1}{\sqrt{qz}} e^{i q z}.
\end{equation}
%It is thus at the locations $ qz \gg \nu$ that the confinement of AdS space drops out.
%and correspondingly for this region of parameters the integrand in $K_\omega$ for AdS (\ref{eq:KomegaAdS}) becomes similar to that in $K_{\omega}$ for Minkowski space (\ref{eq:Komega}). The $z$ component of the modes relates its value at finite $z$ to the value at the boundary.
%In Minkowski space, this relation was $\exp(i \sqrt{\omega^2 - k_x^2} \ z)$. In AdS, for $ q z \gg \nu$, \ref{eq:J} tell us the relation is identical (aside from the irrelevant universal factor of $z^{\Delta}$):$\ z^{\Delta} \exp(i \sqrt{\omega^2 - k_x^2})$. We therefore expect to find similar results in AdS as those in Minkowski space.

It now remains to understand when the radial $z$-component of the momentum, $q$, is real-valued.
In pure AdS space,  the effective potential is bounded from below  $V(z) \ge {\bf k}_x^2$, so the frequency is bounded by
$\omega^2 \ge {\bf k}_x^2$. This translates into the statement that $q^2 = \omega^2 - {\bf k}_x^2\ge 0$.
We find that, as for homogeneous Minkowski space,  pure AdS space has only propagating modes.
If the bulk space deviates from the AdS space, so long as the effective potential $V(z)> {\bf k}_x^2$ for all $z$,
there can only exist propagating modes. With propagating modes only,
the integrals that define the smearing function (\ref{eq:KomegaAdS}) and (\ref{eq:KKomega}) are convergent
and the reconstruction is perfect.
We conclude the the simple criteria for exact reconstruction of the bulk near the boundary is that there are only propagating modes.
%$k_x$ is required to be less than $\omega$. In pure  AdS, one has the same restriction, as a

%\begin{comment}
%XXX PUT SOMEWHERE
At this point, we point out the need to exercise caution.
While the reconstruction for the near-boundary region appears independent of the geometry deep in the bulk,
this is actually incorrect. For instance, as we will see below, the presence of a black hole deep in the bulk drastically modifies
our ability to reconstruct the field anywhere in the bulk.
This is a consequence of the black hole changing the boundary conditions deep in the bulk and permitting evanescent modes
to appear.
%\end{comment}

\subsubsection*{Evanescent modes }
Let us now assume that there is a change in the geometry deep inside the bulk such that the effective potential dips to
$V(z) < {\bf k}_x^2$ and evanescent modes can occur.
For instance, if there is a black brane in the bulk of the Poincar\'e patch AdS space, then the effective potential
$V(z)$ will vanish at the horizon and a continuum of evanescent modes will be present just outside the black brane horizon.
We will discuss in section~\ref{sec:genBack} if there are other geometries for which evanescent modes can occur.
For now, we simply assume that evanescent modes are present and study their impact on the reconstruction.

Choosing $|{\bf k}_x|> \omega$ leads the $z$-component of the modes (\ref{eq:AdSModes}) to become
\begin{equation}
J_{\nu}(q z) = J_{\nu}(i \kappa_z z) = e^{i \nu \pi/2} I_{\nu}(\kappa_z z),
\end{equation}
where $\kappa_z \equiv \sqrt{{\bf k}_x^2- \omega^2}$. For large $\kappa_z z$,
\begin{equation} \label{eq:expI}
I_{\nu}(\kappa_z z) \sim e^{\kappa_z z} ,
\end{equation}
and so this solution is precisely an evanescent mode. It is clear that the monochromatic smearing function (\ref{eq:KomegaAdS}) is divergent if evanescent modes are included,
and we can no longer exactly reconstruct the field. This situation is completely analogous to what happened in the context
of Minkowski space in section~\ref{sec:eva}.

The reader may find it confusing that we can not reconstruct a local field even
close to the boundary, where the bulk asymptotes to pure AdS space. Indeed, the AdS/CFT dictionary (\ref{eq:Ext}) tells us
that knowledge of the local CFT operator $O({\bf x},t)$ should give the local bulk field $\phi({\bf x},t,z)$ at the boundary $z\rightarrow 0$.
If we were to regulate this expression at $z=\epsilon$ so as to say that $O({\bf x},t) = \epsilon^{- \Delta} \phi({\bf x},t,\epsilon)$, then we would appear to have a contradiction. The resolution is that the proximity to the boundary,
in the sense of the $z$-location at which modes encounter the confining barrier of the AdS space,
depends on the momentum of the mode. We see from (\ref{eq:J}) that the mode at $z \gg \nu q^{-1}$ behaves as if
it is in flat space, while the mode at $z \lesssim \nu q^{-1}$ experiences the AdS confining barrier and decays as $z^{\Delta}$.
It is the latter region for which (\ref{eq:Ext}) applies. In other words, the $\epsilon$ for which (\ref{eq:Ext}) starts to become
applicable is momentum-dependent.
%In pure AdS it is forbidden since it diverges as $z\rightarrow \infty$. However, if there is a sufficient change in the geometry at large $z$, then these modes may become allowed.

\subsubsection*{AdS locality}
If the evanescent modes are present, the spectral integral for the smearing function (\ref{eq:KomegaAdS}) is divergent
and we can not reconstruct $\phi({\bf x},t,z)$. The best we can hope for is to reconstruct a coarse-grained bulk field.
As in the Minkowski case, we would like to reconstruct the field coarse-grained in the ${\bf x}$-direction with resolution
scale $\sigma$ (\ref{eq:phiSigma}). Thus, we should compute the $\sigma$-grained smearing function
\bea
K^\sigma_\omega({\bf x}, z|{\bf x}') = \int_b \rmd \bar{\bf x} \ e^{-({\bf x} - \bar{\bf x})^2/\sigma^2} \ K_\omega(\bar{\bf x}, z | {\bf x}') \ ,
\eea
where $K_{\omega}$ appearing in the integrand is given by (\ref{eq:KomegaAdS}). Hence,
\begin{equation} \label{eq:KregAdS}
K_{\omega}^{\sigma} ({\bf x},z| {\bf x}') = 2^{\nu} \Gamma(\nu+1) \int_{-\infty}^\infty {\rmd {\bf k}_x\ e^{- i {\bf k}_x \cdot
({\bf x}-{\bf x}')}\ z^{d/2} q^{-\nu}  J_{\nu} \left(\sqrt{\omega^2 - {\bf k}_x^2}\ z \right)\ e^{- {\bf k}_x^2 \sigma^2}} \ .
\end{equation}
In the evanescent regime ($|{\bf k}_x| >\omega$) and at bulk radial position $z  \gtrsim \nu / \sqrt{{\bf k}_x^2 - \omega^2}$,
the Bessel function undergoes exponential growth (\ref{eq:expI}).
Since $\nu$ is assumed to be of order $1$, evaluating (\ref{eq:KregAdS}) leads to the same behavior as
in the analogous expression for the Minkowski case (\ref{eq:Exp}).
Thus, we conclude that we can only reconstruct the $\sigma$-grained field in the bulk to the depth
\begin{equation} \label{eq:sigmaz}
\sigma \gtrsim z ~.
\end{equation}
Going to a bulk coordinate distance $z$ deeper than $\sigma$ would require exponential precision
in the measurements of $\phi_0({\bf x},t)$ at the timelike boundary hypersurface ($z=0$).

So far, the conclusions are qualitatively the same as for the Minkowski space problem. In AdS space, we need to go one step further and
express the resolution bounds in proper distances. The distance $\sigma$ in which we have coarse-grained in the $x$-direction is a coordinate distance, natural from the viewpoint of boundary data of the dual CFT.
The proper distance between two points $(z_1, {\bf x}_1, t_1)$ and $(z_2, {\bf x}_2, t_2)$ in AdS space is given by
\bea
(\Delta s)^2 = {1 \over z_1 z_2} [ (z_1 - z_2)^2 + ({\bf x}_1 - {\bf x}_2)^2 - (t_1 - t_2)^2]~.
\eea
So, converting the coordinate distance resolution $|\Delta {\bf x}| \ge \sigma$ to the proper one,
we have the proper coarse-graining scale at bulk location $z$
\begin{equation}
\sigma_{\rm proper}(z) = \sigma \frac{L_{\rm AdS}}{z}.
\end{equation}
Thus, (\ref{eq:sigmaz}) translates at radial depth $z$ to a proper resolution bound:
\begin{equation} \label{eq:sigmaz2}
\sigma_{\rm proper}(z) \gtrsim L_{\rm AdS} .
\end{equation}
This is one of our main results:  with our assumptions,
{\sl in an asymptotically AdS space which gives rise to evanescent modes, we can not reconstruct the bulk field any better than the AdS scale.}
Moreover, the right-hand side of (\ref{eq:sigmaz2}) being independent of $z$, we are able to
reconstruct the AdS bulk without limits on the depth, at least for the near-boundary region.\footnote{The result (\ref{eq:sigmaz2}) was  anticipated  in \cite{BouFre12} in the context of  AdS-Rindler.}

We derived the proper resolution bound (\ref{eq:sigmaz2}) from  analysis of the near-boundary region of the bulk.
For regions deeper in, the bound may be further modified. For instance, for AdS black holes,
the bound could also depend on the black hole horizon scale $R_{\rm BH}$ as well as the bulk depth $z$.
On general grounds, we expect that the resolution bound takes the form
\begin{equation}\label{correction}
\sigma_{\rm proper}(z) \gtrsim L_{\rm AdS} \ \left[ 1 + \varepsilon (z, R_{\rm BH}) \right],
\end{equation}
where $\varepsilon$ is background-specific correction factor at the bulk depth $z$.
In the next section, we shall extract this correction for a large AdS black hole.

Once again, we have a clear analogy with Scanning Tunneling Microscopy (STM).
The invention of STM revolutionized microscopy: by placing a needle at a distance $\sigma$ from a sample,
some of the evanescent modes could be captured, allowing resolution on a scale $\sigma$.
Remarkably, in AdS space, the resolution in the bulk is set by the AdS scale and
the reconstruction can be done to arbitrary bulk depth.

\subsection{Bulk reconstruction deeper in} \label{sec:genBack}
In this section,  we evolve deeper into the bulk. Specifically, we take an AdS-Schwarzschild background and
classify the types of modes present, and identify criteria for the presence of evanescent modes in other backgrounds.
We also find the correction $\varepsilon$ in (\ref{correction}) that the resolution bound (\ref{eq:sigmaz2}) receives at locations deeper in the bulk.

\subsubsection*{Effective potential for modes}
Consider a scalar field $\phi$ in a general spherically symmetric background that asymptotes to AdS space,
\begin{equation} \label{eq:Static}
\rmd s^2 = - f(r) \rmd t^2 + \frac{\rmd r^2}{f(r)} + r^2 \rmd\Omega_{d-1}^2.
\end{equation}
The field $\phi$ satisfies the wave equation (\ref{eq:wave}). Separating $\phi$ as
\begin{equation} \label{eq:phi}
\phi(r,t,\Omega) = \varphi(r) Y(\Omega) e^{- i \omega t}
\end{equation}
gives for the radial field $\varphi(r)$,
\begin{equation} \label{eq:rad1}
\frac{\omega^2}{f} \varphi + \frac{1}{r^{d-1}}\partial_r (f\ r^{d-1} \partial_{r} \varphi) - \frac{l(l+d-2)}{r^2}\varphi - m^2 \varphi = 0.
\end{equation}
Letting $\varphi(r) = u(r)/r^{\frac{d-1}{2}}$ and changing variables to the tortoise coordinate $dr_{*} = f^{-1} dr$ turns (\ref{eq:rad1}) into a Schr\"odinger-like equation
\begin{equation} \label{eq:Schro}
\frac{d^2 u}{dr_{*}^2} + (\omega^2 - V(r))u=0,
\end{equation}
with the effective potential
\begin{equation} \label{eq:Pot}
V(r)= f\left[ \frac{(d-1)}{2} \frac{f'}{r} + \frac{(d-1)(d-3)}{4} \frac{f}{r^2} + \frac{l(l+d-2)}{r^2} +m^2 \right] .
\end{equation}

\subsubsection*{Small AdS-Schwarzschild Black Hole}

\begin{figure}[tbp]
\centering
\subfigure[]{
	\includegraphics[width=2.7in]{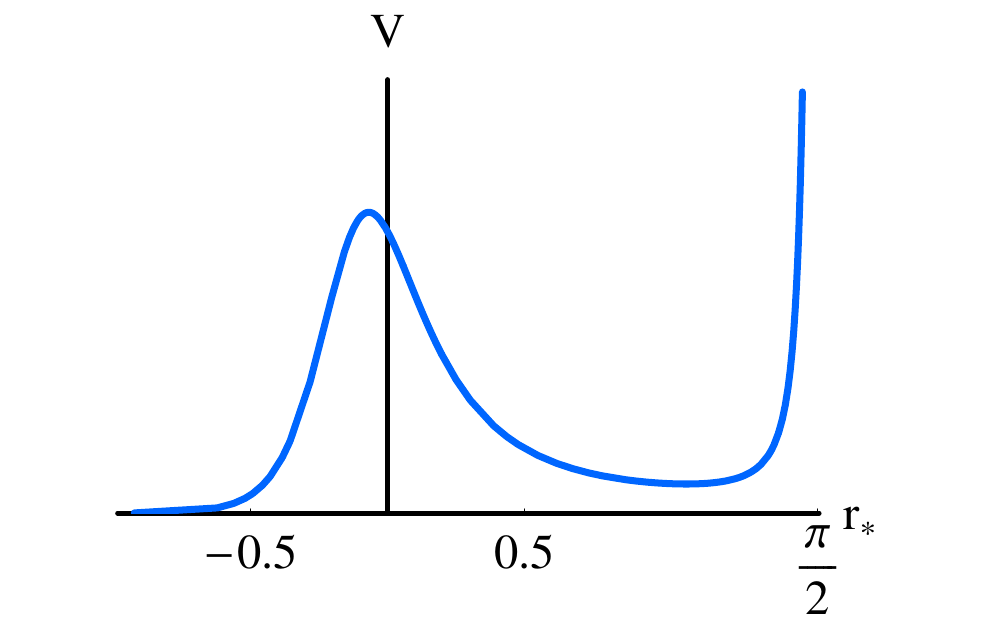}
	}
	\hspace{.2in}
		\subfigure[]{
	\includegraphics[width=2.7in]{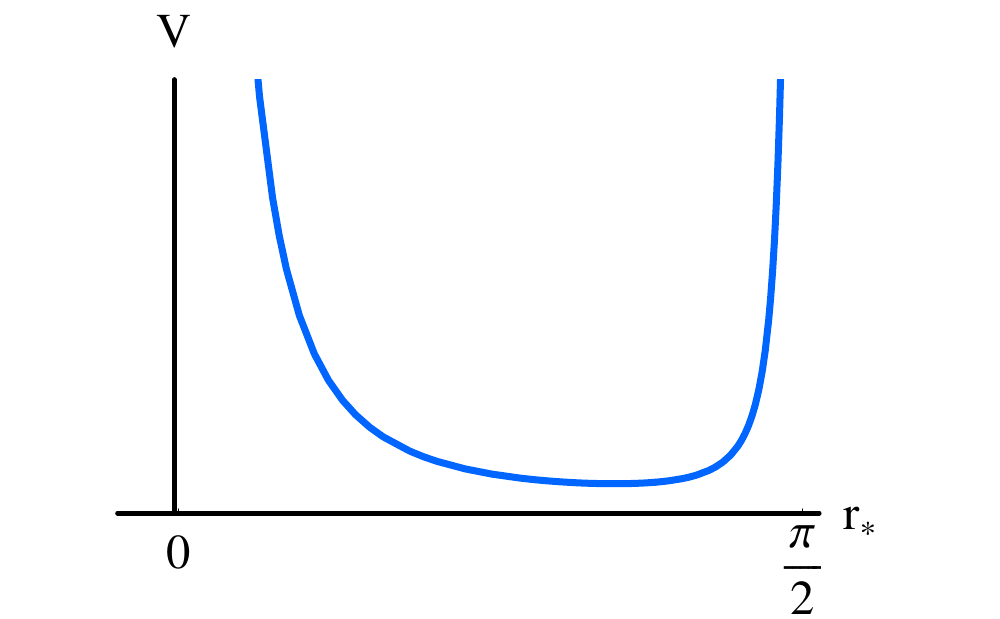}}
\caption{
\sl (a) The effective potential (\ref{eq:Pot}) for a small AdS black hole. The boundary of AdS space is at tortoise coordinate $r_* =\pi/2$, while the horizon is at $r_* = - \infty$. Evanescent modes arise if the potential ever drops below the value at the local minimum present in the pure AdS space ($\sim l^2$). Here, this occurs because the potential approaches zero at the horizon. (b) The effective potential for pure global AdS space.}  \label{fig:smallBH}
\end{figure}

The AdS- Schwarzschild black hole metric is given by:
\begin{equation} \label{eq:Sch}
\rmd s^2 = - \left(1+ r^2 - \left(\frac{r_0}{r}\right)^{d-2}\right)dt^2 + \frac{dr^2}{1+ r^2 - \left(\frac{r_0}{r}\right)^{d-2}} + r^2 d\Omega_{d-1}^2 .
\end{equation}
Here, we will be interested in a small black hole, so $r_0 \ll L_{AdS}$ and $r_0 \approx 2 M$. In Fig.~\ref{fig:smallBH}(a), we plotted the effective potential (\ref{eq:Pot}) for a small black hole. There are four kinds of modes\footnote{One should note that we are solving for the normal modes since we want a complete basis of solutions to the wave equation. The only boundary conditions we are imposing are that the normal modes be normalizable. In particular, we are \textit{not} imposing infalling boundary conditions at the horizon (we are not solving for quasinormal modes).}, and can be classified depending on the ratio of $\omega$ to $l$. We characterize the modes going from high $\omega$ to low $\omega$.
\begin{enumerate}
\item $\omega \gtrsim  \frac{l}{r_0}$. These modes are higher than the angular momentum barrier and propagate into the black hole. They correspond to throwing $\phi$ particles into the black hole.
\item $ l\lesssim \omega \lesssim  \frac{l}{r_0}$. These modes are directly related to the modes in the pure AdS space (plotted in Fig.~\ref{fig:smallBH} (b)). They correspond to the particles having sufficient angular momentum so that they stay far away from the black hole and do not notice its presence. They do differ from the pure AdS modes in that they have exponentially suppressed tails which are propagating near the black hole.
\item $ l\lesssim \omega \lesssim   \frac{l}{r_0}$. This is the same regime as type 2 modes, but these modes are the ones that have most of their support close to the black hole and only an exponentially small amplitude in the asymptotic region. We will call these modes trapped modes.
\item $\omega \lesssim l$. These are the evanescent modes. All possible evanescent modes, with any $\omega$ for any value of $l$, are present as a result of the potential dropping to zero at the horizon.
\end{enumerate}
As discussed in Sec.~\ref{sec:nearBdy}, it is only the evanescent modes which inhibit reconstruction of the region near the boundary. The trapped modes inhibit reconstruction of the field, but only for regions close to the black hole ($r< 3 r_0/2$). Recall that we normalize all the modes so that their boundary limit is $\phi_{\omega l}r^{\Delta} \rightarrow 1$. The trapped modes are propagating  in most of the AdS bulk. Only when they encounter $r \sim 3 r_0/2$, and have to pass under the centrifugal barrier, do these modes begin to undergo exponential growth. Neither modes of type (1) or (2) are problematic for reconstruction as they never undergo exponential growth as compared to their boundary value.

We should emphasize that the difficulties with reconstruction in a black hole background are not due to the presence of the  horizon, per se. One might suppose there should be difficulties with reconstruction because there are now things which fall into the black hole and so can't be seen from the boundary (in other words, that modes of type (1) cause difficulties). However, the case of pure Poincar\'e Patch provides a counter-example. There, everything falls into the Poincar\'e horizon, yet as we showed in section~\ref{sec:nearBdy}, there are no difficulties with bulk reconstruction. The only necessary criterion the modes need to satisfy for the bulk reconstruction is that their magnitude at the bulk point of interest is not exponentially larger than their boundary value.

It is interesting to ask in which backgrounds, aside from the AdS black hole, the trapped modes (type 3) are present. For a static, spherically symmetric spacetime,
\begin{equation}
\rmd s^2 = - f(r) dt^2 +\frac{dr^2}{g(r)} + r^2 d\Omega^2,
\end{equation}
one could write down an effective potential (similar to (\ref{eq:Pot})) and see if it has a barrier (a local maximum). For the question of reconstruction, we are most interested in modes with large $l$: if there are trapped/evanescent modes it is these that will be hardest to reconstruct. So, we can simplify the analysis by taking the large $l$ limit of the effective potential. For any given $r$, in the large $l$ limit, the effective potential simplifies to
\begin{equation} \label{eq:V}
V = f \frac{l^2}{r^2} ,
\end{equation}
and the criterion for having trapped modes is that
\begin{equation} \label{eq:noSmear}
\frac{\rmd}{\rmd r}\left(\frac{f}{r^2}\right) >0
\end{equation}
for some value of $r$. In \cite{Leich13},  it was shown that (\ref{eq:noSmear}) implies that there is no smearing function for some region in the bulk. The condition (\ref{eq:noSmear}) can be achieved by having a sufficiently dense planet so that its radius is less than $3 M$.

It is also interesting to ask in which backgrounds, aside from the AdS black hole, are the evanescent modes (type 4) present. The criterion for having evanescent modes is that the effective potential has a new global minimum, different from the one present in pure AdS space which has the value $\sim l^2$. From (\ref{eq:V}), this means that
\begin{equation} \label{eq:fcond}
\frac{f}{r^2} < 1
\end{equation}
for some value of $r$. Outside the planet, $r>R$, the metric takes the form (\ref{eq:Sch}). Letting $r_h$ denote the would-be horizon of the planet, (\ref{eq:fcond}) translates into the following condition on the radius $R$ of the planet:
\begin{equation}
R - r_h < \frac{r_h}{d-2} \left(\frac{r_h}{L_{AdS}}\right)^2 ~.
\end{equation}
Since $r_h/L_{AdS} \ll 1$ by assumption, it is nontrivial for physical matter to be so dense.  In particular, if one assumes the density is nonnegative and a monotone decreasing function, then a general argument \cite{Wald} asserts that one must have $R- r_h > r_h/8$ (for $4$ spacetime dimensions).

\subsubsection*{Large AdS-Schwarzschild black hole}

\begin{figure}[tbp]
\centering
	\includegraphics[width=3in]{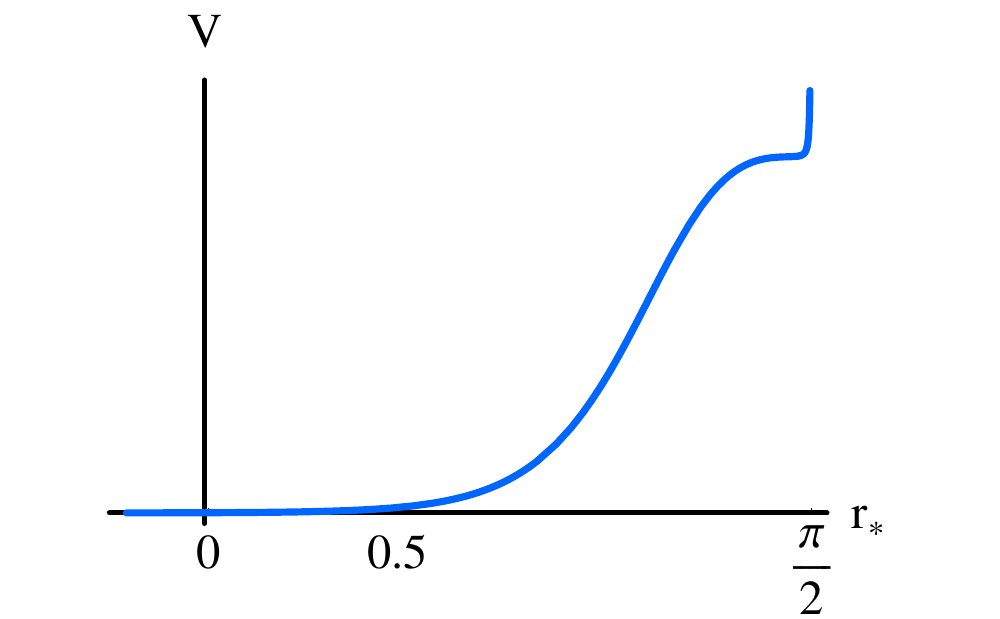}
\caption{\sl The effective potential for a large black as a function of the tortoise coordinate. Note that, with tortoise coordinates, the region near the boundary (large $r$) gets compressed near $\pi/2$. Thus, the top corner of the plot is the only portion of the potential that is similar to a portion of the pure AdS effective potential (Fig.~\ref{fig:smallBH}(b)). Here, the modes are mostly those of type (1) that propagate into the black hole horizon, and type (4) that have low energy and hence are evanescent.  } \label{fig:largeBH}
\end{figure}

We would like to find corrections to the reconstruction resolution bound $\sigma_{\text{proper}}>L_{AdS}$ for locations deeper in the bulk, where the geometry is no longer pure AdS space. To find this, we need to solve for the radial profile of the evanescent modes. The evanescent modes will grow exponentially as one moves from the boundary into the bulk. The distance into the bulk that we can evolve is set by the rate of the growth: when the coefficient in the exponential becomes of order one, we are unable to evolve deeper in.

We focus on a large AdS black hole, which is similar to a black brane. In the limit that the black hole is large, $r_0\gg 1$,  the metric (\ref{eq:Sch}) can be simplified by dropping the $1$ as it is small relative to $r^2$, so that $f(r) = r^2-(r_0/r)^{d-2}$. If in addition to this we also zoom in on a small portion of the solid angle, this yields the black brane metric,
\begin{equation} \label{eq:Bbrane}
\rmd s^2 = - \left(r^2 - \left(\frac{r_0}{r}\right)^{d-2}\right)\rmd t^2 + \frac{\rmd r^2}{ r^2 - \left(\frac{r_0}{r}\right)^{d-2}} + r^2 \rmd {\bf x}^2 .
\end{equation}
The horizon location, $r_h$, and temperature, $T$, are given by
\begin{equation}
r_h^d = r_0^{d-2} \ , \ \ \ \ \ \ \ \  T= \frac{d\ r_h}{4 \pi} ~.
\end{equation}
The equation for the effective potential satisfied by the radial modes is given by (\ref{eq:Pot}) with $l(l+d-2)$ replaced by $k_x^2$,
\begin{equation}
V= f\left[\frac{d^2-1}{4} +m^2 + \frac{(d-1)^2}{4} \frac{r_0^{d-2}}{r^d}  +\frac{k_x^2}{r^2}\right]~.
\end{equation}
To solve for the radial profile of the modes,  we use the WKB approximation, only keeping the exponential factor
\begin{equation} \label{eq:u1}
u(r) = \exp\left(\int_{r}^{\infty}{\frac{\rmd r'}{f(r')}}\sqrt{V(r') - \omega^2} \right) ~.
\end{equation}
We are interested in the mode which is the most evanescent, and this is the one with $\omega=0$. Thus, for (\ref{eq:u1}), we have
\begin{equation}\label{eq:u2}
u(r) = \exp\left(\int_r^{\infty}{\frac{\rmd r'}{\sqrt{r'^2 - r_h^d/r'^{d-2}}}\sqrt{\bar{\nu}^2 +\frac{k_x^2}{r'^2} +\frac{(d-1)^2}{4}\left(\frac{r_h}{r'}\right)^d }}\right) ~,
\end{equation}
where we defined $\bar{\nu}^2 = m^2 + (d^2-1)/4$. Since $r_h/r < 1$ and $\bar{\nu}$ is of order $1$, we can drop the last term in (\ref{eq:u2}). Converting to the $z$-coordinate, $z=1/r$, and setting $\sigma \equiv k^{-1}$, (\ref{eq:u2}) becomes
\begin{equation} \label{eq:u3}
u(z) = \exp\left(\int_0^{z}{\frac{\rmd z'}{\sqrt{1-(z'/z_h)^d}}\sqrt{\frac{1}{\sigma^2}+\frac{\bar{\nu}^2}{z'^2}}}\right) ~.
\end{equation}
In the case of pure AdS space, we have (\ref{eq:u3}) but with $z_h= \infty$. In that case, we argued that for small $z$ (such that $z \ll \sigma$), we can drop the $1/\sigma^2$ term in the square-root and find that $u(z)$ has power-law behavior consistent with the conformal scaling, $z^{\Delta}$, of the field near the boundary. For $z \gtrsim \sigma$, we instead drop the $\bar{\nu}^2/z^2$ term in the square root, leading to the behavior
\begin{equation}
u(z) = \exp\left(\frac{z}{\sigma}\right) ~.
\end{equation}
With an AdS black hole present, this behavior receives only minor corrections. Since
\begin{equation}
\frac{1}{\sqrt{1 - (z/z_h)^d}} >1 ~,
\end{equation}
$u(z)$ has a faster growth in an AdS black hole background than in pure AdS space. We can find the field at $z=z_h$ and in the limit $\sigma \ll z_h$.
This gives\footnote{Here, we set the lower limit of integration to be $\sigma$ since this is around where the approximation of dropping the $\bar{\nu}^2/z^2$ term becomes invalid. For $z\lesssim \sigma$, one should instead drop the $1/\sigma^2$ term. However, since by assumption $\sigma/z_h \ll 1$, one can just as well set the lower limit of integration to $0$.},
\begin{equation} \label{eq:u4}
u(z_h) = \exp\left(\frac{1}{\sigma}\int_{\sigma}^{z_h}{\frac{\rmd z}{\sqrt{1 - (z/z_h)^d}}}\right) ~.
\end{equation}
Evaluating the integral gives
\begin{equation} \label{eq:uzh}
u(z_h) = \exp\left( \alpha\ \frac{z_h}{\sigma}\right) ~,
\end{equation}
where $\alpha$ is the scaling exponent
\begin{equation} \label{eq:alpha}
\alpha = \frac{ \sqrt{\pi}\ \Gamma[1+ 1/d]}{\Gamma[2 + 1/d]} ~.
\end{equation}
The value of the scaling exponent $\alpha$ is larger than 1: for instance, in $d=4$, $\alpha \approx 1.3$, only slightly larger than $1$. This fits with our expectation since,  if we had been in pure AdS space, then we would have had
$u(z) = \exp(z/\sigma)$. The scaling exponent $\alpha$ is universal in that it depends only on $d$ but not on other details of the AdS black hole; it can be considered a nontrivial prediction for a strongly coupled CFT at finite temperature.

Comparing with (\ref{correction}), we now obtain the correction factor $\varepsilon$ to the resolution bound at the horizon:
\begin{equation}
\varepsilon (z_h) = \alpha - 1 = \frac{ \sqrt{\pi}\ \Gamma[1+ 1/d]}{\Gamma[2 + 1/d]} - 1 \le 0.7\cdots.
\end{equation}
%It is interesting that the correction is actually insensitive to the size of black hole horizon $R_{\rm BH}$ and that it is the largest for $d \rightarrow \infty$.
In the next section, we will find physical interpretation of the correction factor $\varepsilon$ at the horizon as a
nonperturbative effect in the thermal Green's functions in the CFT dual.

\section{Evanescence in CFT Dual} \label{sec:CFT}
In this section, we turn to the CFT dual of the AdS space and pose the question: how do evanescent modes manifest themselves in the CFT?
The first question is whether the evanescent modes are part of the CFT spectrum. To be specific, consider the large AdS black hole we studied in the last section. The black hole is the holographic dual of the CFT at a high temperature $T$ that equals the black hole's Hawking temperature. The $T=0$ vacuum is now changed to the $T>0$ ground state (this ground state is created out of the vacuum by a micro-canonical operator ${\cal O}_{\rm BH}$). Now, the $T>0$ spectrum includes the continuous, evanescent spectrum in addition to the discrete spectrum. In effect, the continuous spectrum increases the dimension of the Hilbert space.

One might object that the evanescent states do not propagate to the AdS boundary and need not be included.
However, they do contribute to CFT correlators at finite temperature and should be included. For instance,
consider the 4-point CFT correlator $G_4$, as computed holographically in a large AdS black hole background:
\bea
G_4 (1, 2, 3,4) &\equiv& \Big\langle T \Big\vert {\cal O}_1 {\cal O}_2 {\cal O}_3{\cal O}_4 \Big\vert T \Big\rangle_{\rm CFT} \nn
&=& \int \!\!\! \int_B \rmd x \rmd y\ G_{Bb}(1, x) G_{Bb} (2, x) K_2(x, y) G_{Bb}(3, y) G_{Bb} (4, y) ~.
\eea
The point is that the integrals over the bulk $x$ and $y$ in the region close to the black hole horizon contain
contributions from the evanescent modes. This is because the evanescent modes are genuine propagating modes in this region
and the bulk-to-bulk propagator $K_2(x, y)$ contains these modes. More generally, propagating modes from or to
the AdS boundary would couple to the evanescent modes whose wave function is localized near the black hole horizon.
This implies that, in the dual CFT, we should expect a new class of 3-point correlators that involve
two local CFT operators ${\cal O}_1, {\cal O}_2$ and one effective micro-canonical operator $E_T$ associated with the heat bath:
\bea\label{eva}
\langle  {\cal O}_1 {\cal O}_2  E_T \rangle  \sim \langle T \vert {\cal O}_1 {\cal O}_2 \vert T \rangle.
\eea
%
%Include optics result
To see (\ref{eva}) explicitly, we compute the CFT 2-point function in the evanescent regime by computing
a bulk 2-point function and taking its boundary limit.
Let us canonically quantize the bulk scalar field $\phi ({\bf x},t,z)$ and expand in terms of modes $\phi_{\omega, {\bf k}}$ ,
\begin{equation}
\phi = \int\!\!\!\int {\rmd \omega \rmd {\bf k} \left( \phi_{\omega, {\bf k} }\ a_{\omega {\bf k}} + \phi_{\omega {\bf k} }^{*}\ a_{\omega {\bf k}}^{\dagger}\right)} ~,
\end{equation}
where the creation operators satisfy the usual commutation relation. Here, we normalize the modes differently from section ~\ref{sec:AdS} and use the bulk Klein-Gordon normalization, which is more natural in the present context. ~\footnote{ In section~\ref{sec:AdS}, for instance in (\ref{eq:AdSModes}),
we normalized the modes so that, when rescaled by $z^{-{\Delta}}$, their $z$-component approaches $1$ at the boundary.
This was a natural normalization in that context since we needed to use
the boundary limit of the field to do a Fourier transform over space and time on the boundary
and extract the spectral coefficients of each modes.
Here, we normalize with respect to the bulk Klein-Gordon norm so that (\ref{eq:bulkTwoPt}) takes a simple form.}

The bulk 2-point function can thus be written as: \footnote{Note that we are taking the bulk state to be the Hartle-Hawking vacuum.}
\begin{equation} \label{eq:bulkTwoPt}
\langle \phi({\bf x}_2,t_2,z_2) \phi({\bf x}_1,t_1,z_1)\rangle = \int {\rmd \omega\ \rmd {\bf k}\ \phi_{\omega {\bf k}}({\bf x}_2,t_2,z_2)\ \phi_{\omega {\bf k}}^*({\bf x}_1,t_1,z_1)}
\end{equation}
Writing the modes as
\begin{equation}
\phi_{\omega {\bf k}}({\bf x},t,z) = f_{\omega {\bf k}}(z) e^{-i {\bf k} \cdot {\bf x} + i \omega t} ~,
\end{equation}
and using the AdS/CFT dictionary (\ref{eq:Ext}), we get
\begin{equation} \label{eq:twoPoint}
\Big\langle T \Big\vert O({\bf x}_2,t_2) O({\bf x}_1,t_1) \Big\vert T \Big\rangle
= \text{lim}_{z\rightarrow 0} \int\!\!\!\int{\rmd \omega\ \rmd {\bf k} \ z^{-2 \Delta} |f_{\omega {\bf k}}(z)|^2 e^{-i {\bf k} \cdot ({\bf x}_2-{\bf x}_1) + i \omega (t_2 - t_1)}} .
\end{equation}
Let us focus on the portion of the 2-point function (\ref{eq:twoPoint}) coming from the evanescent regime $\omega \ll k$.
Since the Klein-Gordon normalization roughly corresponds to having the mode of order $1$ at the horizon,
to find the $z\rightarrow 0$ limit of $|f_{\omega {\bf k}}(z)|^2$, we just need the WKB factor giving the relative suppression
at the boundary. This factor was computed in (\ref{eq:uzh}). We thus find that,
in the regime $\omega \ll |{\bf k}|$ and $|{\bf k}|\gg T$, the bulk 2-point function has the behavior (ignoring pre-factors),
\begin{equation} \label{eq:twopointF}
G(\omega, {\bf k} ) \simeq e^{- \tilde{\alpha} \frac{|{\bf k}|}{T}} \qquad \mbox{where} \qquad \tilde{\alpha} = {d \over 2 \pi} \alpha \ .
\end{equation}
Here, $\tilde{\alpha}$ is a constant factor proportional to the scaling exponent $\alpha$ we previously encountered in (\ref{eq:alpha}).\footnote{
In \cite{SonSta02},
the imaginary part of the retarded Green's function was studied in the evanescent regime by an alternate method.
Their result for $d=4$ agrees with our (\ref{eq:twopointF}).}
The result (\ref{eq:twopointF}) is not surprising.
At zero temperature, there is no evanescent mode with $\omega < |{\bf k}|$
 and so such a correction should vanish. So, at finite temperature, the evanescent modes must generate an equilibrium thermodynamic contribution
 which vanishes in the zero temperature limit. This explains the Boltzmann distribution (\ref{eq:twopointF})
 of the evanescent mode contribution.

In the previous sections, we argued that a black hole in the bulk changes the boundary conditions so as to permit evanescent modes and that a small error in determining their coefficient will get exponentially amplified when reconstructing the bulk.
Here, we have just taken the ground state for the large black hole background
and found the boundary imprint of the evanescent modes through CFT 2-point correlators at finite temperature.
If the bulk state is a slight deviation from this ground state, then to determine it near the horizon one must make a boundary measurement precise enough to detect something as small as (\ref{eq:twopointF}).
Of course, we can have an excited state in the black hole background that has a much larger coefficient for an evanescent mode.

\section{Discussion}
The AdS resolution bound we have found in this paper has important impacts on the dual CFT.
The first question concerns the boundary CFT data.
We showed that, in an AdS black hole background, bulk fields have evanescent modes.
These modes are exponentially suppressed near the boundary.
Translated into the CFT description, this means that we are unable to reconstruct the bulk on sub-AdS scales
if we only have local CFT data, $\langle O \rangle$, and do not have it to exponential precision.

The second question is what CFT data naturally encodes the evanescent modes. Much in the same way as the evanescent modes trapped to a material are regarded as ``part'' of the material,
the evanescent modes in an AdS black hole background are naturally part of the semiclassical black hole.
Thus, a speculative possibility is that there exists a theory which describes the near-horizon atmosphere of a black hole
and, as a subsector of the CFT, the theory can be thought of as living on the black hole horizon.~\footnote{Note that the black hole horizon would be used as a holographic screen to describe the black hole atmosphere;
we are not discussing anything about the black hole interior.}

An extremal Reissner-Norstrom AdS black hole is a particularly concrete setting in which to explore this possibility,
as the near horizon limit is AdS$_2 \times \mathbb{S}^{d-1}$. From the perspective of the AdS$_2$,
the evanescent modes of the underlying AdS black hole are propagating modes of AdS$_2$.
Furthermore, AdS$_2 \times \mathbb{S}^{d-1}$ should be dual to multiple copies of a CFT$_1$.
We would therefore conjecture that the bulk of the AdS extremal black hole could be reconstructed by a combination of local operator data coming from both the boundary CFT$_d$ and the tower of horizon CFT$_1$ 's.~\footnote{AdS$_2
\times \mathbb{S}^{d-1}$ has been argued to be problematic \cite{Mal98,Almh14}. This would not be a problem for us, as we do not fully take the near horizon limit.
The complete near horizon limit would correspond to keeping only the $\omega =0$ modes,
while we are interested in small but finite $\omega$ modes.
Alternatively, one could consider our construction for a black brane for which the quoted issues are absent.}

\acknowledgments
We thank Ofer Aharony, Dongsu Bak, Raphael Bousso, Donghyun Cho, Shmuel Elitzur, Ben Freivogel, Daniel Harlow, Nissan Itzhaki, Wonho Jhe, Stefan Leichenauer, Juan Maldacena, Yaron Oz, Joe Polchinski, Douglas Stanford and Herman Verlinde for discussions. %VR is especially grateful to R. Bousso, S. Leichenauer, and B. Freivogel for many discussions and previous collaboration.
The work of SJR is supported in part by the National Research Foundation of Korea funded by the Korea government through grants 2005-0093843, 2010-220-C00003 and 2012K2A1A9055280. The work of VR is supported by the Berkeley Center for Theoretical Physics.

\appendix

\section{Evanescent Optics} \label{sec:EvaOptics}
In this appendix, we summarize some central principles of optics that make use of evanescent modes.
\subsection*{Total Internal Reflection} \label{sec:tir}
\begin{figure}[tbp]
\centering
%\subfigure[]{
	\includegraphics[width=2.5in]{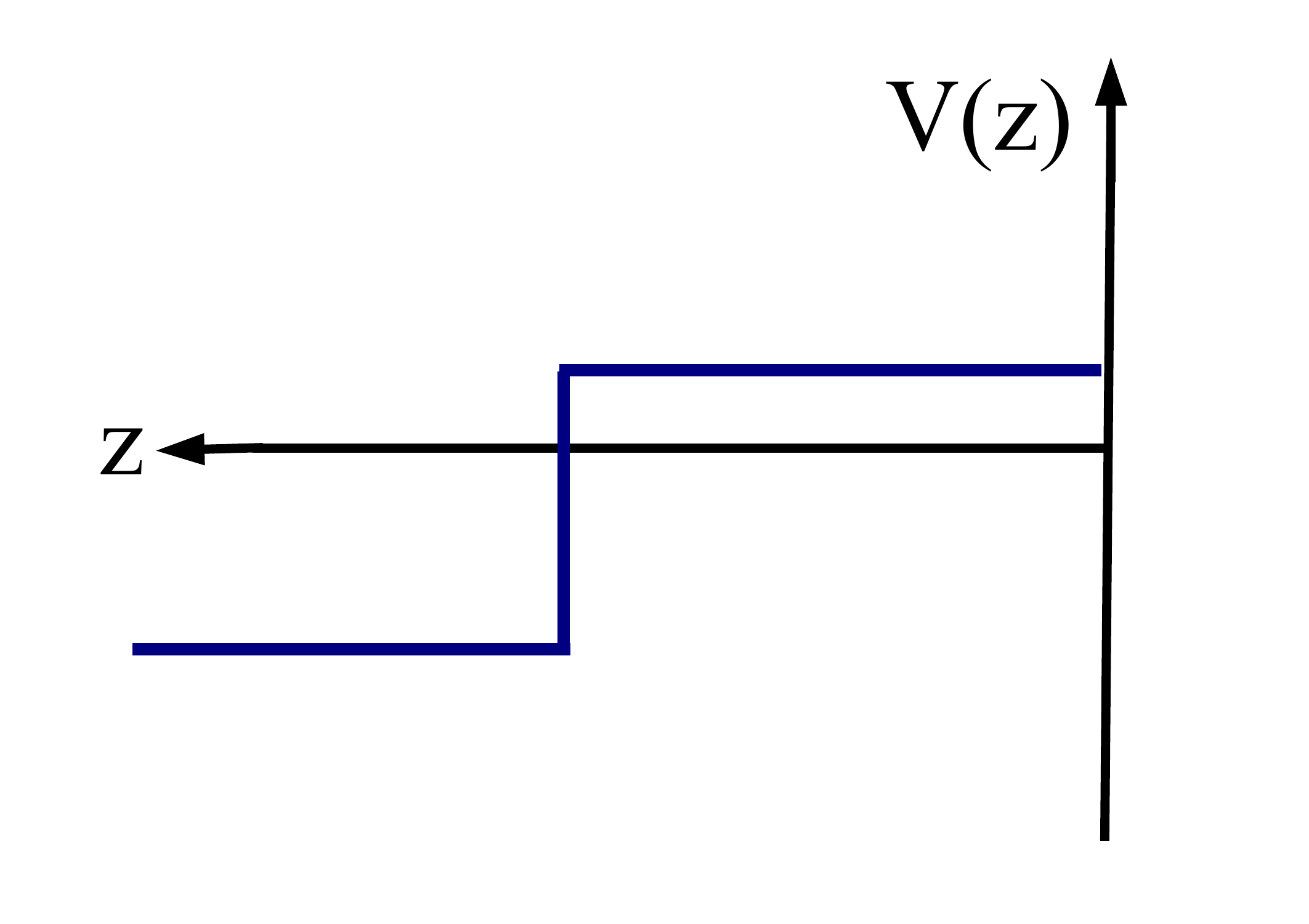}
%	}
\caption{\sl The effective potential for the wave equation in a medium that undergoes a jump in its index of refraction.}
\label{fig:VeffC}
\end{figure}

To see how evanescent waves can be generated if homogeneity is broken, consider the wave equation in a background which has a $z$-dependent speed of light,
\begin{equation}
\left(-\frac{1}{c(z)^2} \partial_t^2 + \partial_x^2 + \partial_z^2\right)\ \phi = 0 .
\end{equation}
The modes are
\begin{equation}
\phi(x,t,z) = e^{- i\omega t - i k_x x}\ u_{\omega, k}(z),
\end{equation}
where $u(z)$ satisfies a Schr\"odinger-like equation
\begin{equation}
-u'' + V(z) u =0
\end{equation}
with an effective potential
\begin{equation} \label{eq:VeffC}
V(z) = k_x^2 - \frac{\omega^2}{c^2(z)}.
\end{equation}
As a simple scenario which generates evanescent modes, we take
\begin{equation}
c(z) = \left\{\begin{array}{rl}
c_1 & \ \ \ z>z_1 \\
c_0 & \ \ \ z\leq z_1 .
\end{array}\right.
\end{equation}
The potential (\ref{eq:VeffC}) for this choice of $c(z)$ is shown in Fig. \ref{fig:VeffC}.
We see that modes with
\[
\frac{\omega^2}{c_1^2}<k_x^2 <\frac{\omega^2}{c_0^2}
\]
are evanescent for $z<z_1$. Indeed, the lower limit $k_x =\omega/c_1$ is precisely the well-known critical angle for total internal refraction: $\sin \theta_c = n_1$. Here, $\theta_c$ is the angle between the incoming wave at $z>z_1$ and the normal in the $z$ direction and $n_1$ is the index of refraction of the material, $n_1=c_0/c_1$.

\subsection*{Microscopes}\label{sec:micro}

\begin{figure}[tbp]
\centering
%\subfigure[]{
	\includegraphics[width=2.5in]{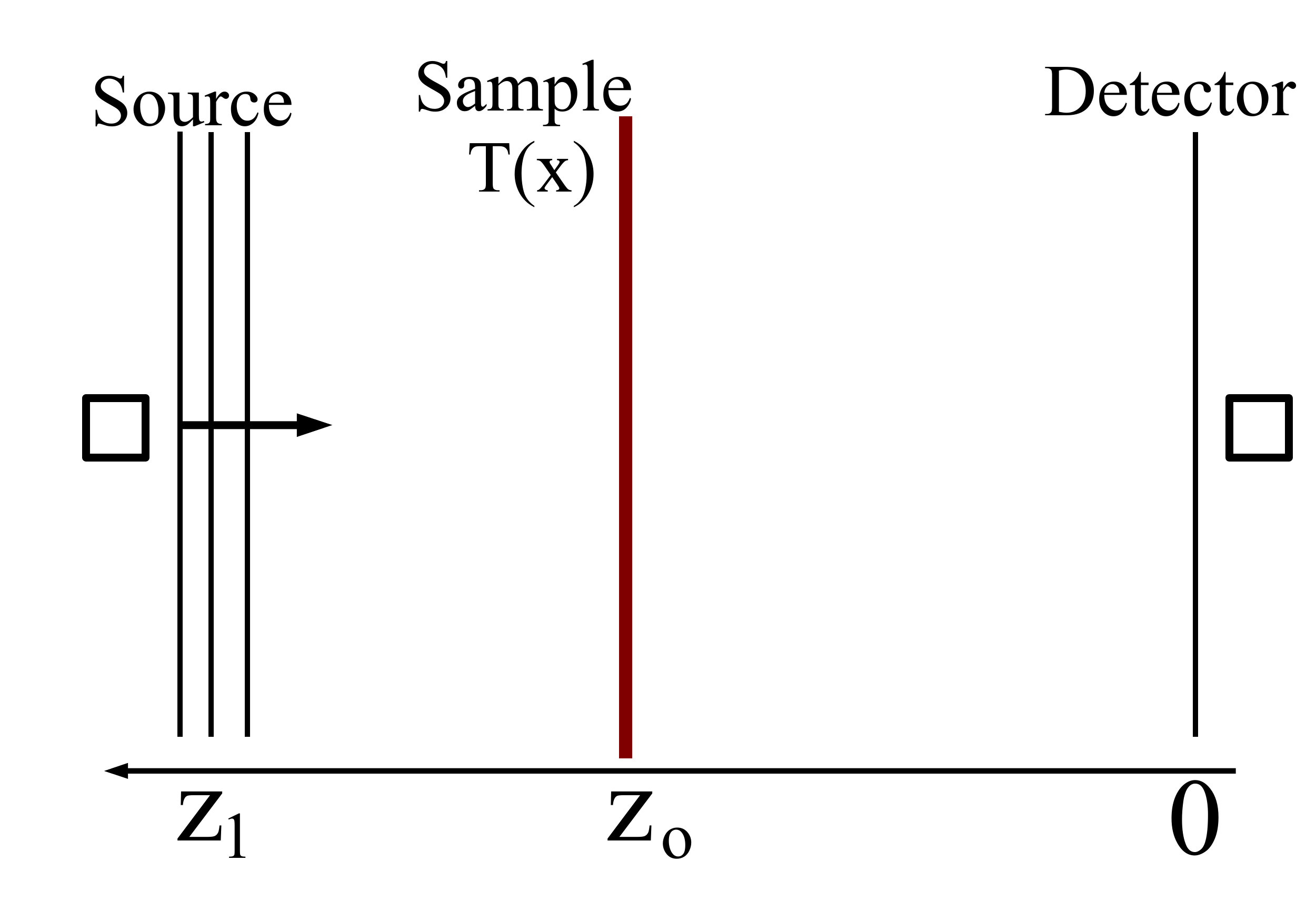}
%	}
\caption{\sl Light is shined from the source ($z_1$), interacts with the sample ($z_0$), and is received at $z=0$. } \label{fig:mic}
\end{figure}
Another context in which evanescent modes appear is in the use of a microscope. The microscope is trying to resolve the spatial features of a sample which is thin and located at $z_0$ (see Fig. \ref{fig:mic}). The sample is projected with monochromatic light of frequency $\omega$ coming from a source at $z_1>z_0$. The sample has some space-dependent transmission coefficient $T(x)$, which then determines the wave profile that is received at the detector at $z=0$. As long as the sample has spatial features on scales smaller than $\omega^{-1}$, evanescent wave which are trapped near the sample will be generated.

% and the goal is to determine $T(x)$ from the wave recieved at $z=0$.
The field consists of a monochromatic wave with frequency $\omega$, $\phi_{\omega}(x,z)$, which we will simply denote by $E(x,z)$. We wish to relate the field at the source to the field received at the detector, and see how the transmission coefficient enters this expression.

The field at the source is denoted by $E_{\text{source}}(x,z_1)$, and the generated wave propagates freely from $z_1$ to $z_0$,
\begin{equation} \label{eq:Ezsz0}
E_{\text{source}}(q_x,z_0) = E_{\text{source}}(q_x,z_1) e^{- i q_z (z_1 -z_0)},
\end{equation}
where $q_z=\sqrt{\omega^2 - q_x^2}$. The effect of the sample is to modify the wave at $z=z_0$ by the transmission coefficient. After passing through the sample, the field becomes
\begin{equation} \label{eq:ET}
E_{\text{sample}}(x,z_0) = T(x) E_{\text{source}}(x,z_0),
\end{equation}
In Fourier space, (\ref{eq:ET}) becomes the convolution
%\begin{equation} \label{eq:Efourier}
%E_{\text{sample}}(k_x,z_0) = \int{dx\ e^{i k_x x}\ E_{\text{sample}}(x,z_0)}.
%\end{equation}
%Inserting (\ref{eq:ET}) into (\ref{eq:Efourier}) gives
\begin{equation} \label{eq:Esamsour}
E_{\text{sample}}(k_x,z_0) = \int{\rmd q_x\ T(k_x - q_x)\ E_{\text{source}}(q_x,z_0)}.
\end{equation}
From the sample, the wave propagates freely to the detector at $z=0$,
\begin{equation} \label{eq:Edetsour}
E_{\text{detector}}(k_x,0) = E_{\text{sample}}(k_z,z_0)\ e^{-i k_z z_0}.
\end{equation}
Combining (\ref{eq:Ezsz0}), (\ref{eq:Esamsour}), and (\ref{eq:Edetsour}) yields
\begin{equation} \label{eq:EFinal}
E_{\text{detector}}(x,0) = \int\!\!\!\int{\rmd k_x\ \rmd q_x\ E_{\text{source}}(q_x,z_1) e^{-i q_z (z_1 -z_0)}\ T(k_x - q_x)\ e^{-i k_z z_0}\ e^{-i k_x x}}.
\end{equation}
The result (\ref{eq:EFinal}) is the relation we had been seeking between the emitted field at the source, $E_{\text{source}}$ at $z=z_1$, and the received field at the detector, $E_{\text{detector}}$ at $z=0$.

From (\ref{eq:EFinal}), we see that the sample serves to convert the waves impinging on it with $x$ momentum $q_x$ to those with momentum $k_x$, with the conversion amplitude given by $T(p_x)$ where $p_x\equiv (k_x-q_x)$. The incident waves on the sample are propagating; thus $|q_x|<\omega$. For the converted waves to be propagating, they need $|k_x|<\omega$. Thus, unless the sample has no features on any spatial scale smaller than $\omega^{-1}$ (so that $T(p_x)$ vanishes for all $|p_x|>\omega$), there will be evanescent modes at $z<z_0$. Indeed, to generate evanescent modes, one only need to choose a sample which is a reflecting metal sheet with a hole. In this case, $T(x)$ has a compact support, requiring $T(p_x)$ to be nonzero at arbitrarily large $p_x$.

\subsection*{Scanning Tunneling Optical Microscope}
We have found that the frequency $\omega$ of the wave shined on the sample sets the limit of the scale on which the sample can be resolved: modes with $k_x>\omega$ are evanescent and too suppressed at the detector location to be measurable. For a long time this was believed to set a fundamental limit on the best achievable resolution of a material \cite{Hecht}.\footnote{It may appear this hindrance can be overcome by simply using waves of arbitrarily high frequency. However, sufficiently high frequency waves damage the sample (via the photoelectric effect).}

\begin{figure}[tbp]
\centering
%\subfigure[]{
	\includegraphics[width=1.4in]{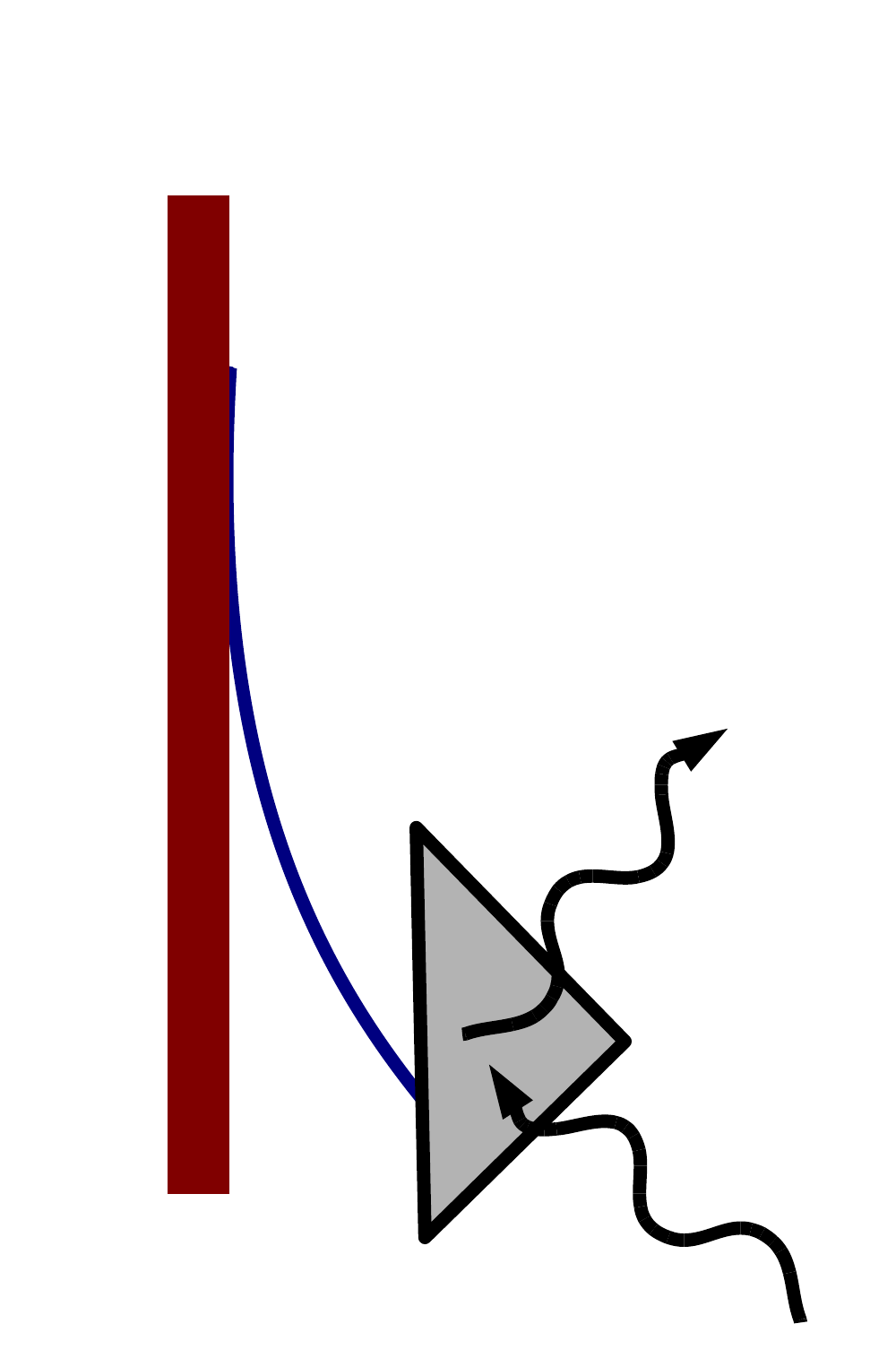}
%	}
\caption{\sl The evanescent modes can be converted into propagating modes by placing, for example, a piece of glass near them.} \label{fig:evaDetect}
\end{figure}
The way to non-invasively resolve the structure on scales shorter than $\omega^{-1}$ would be to detect the evanescent modes. Since their magnitude is exponentially small at the location of the detector, one must instead detect them close to the sample or, equivalently, convert them into propagating waves. This can be achieved by placing some object near the material so as to change the index of refraction. A sketch is shown in Fig.~\ref{fig:evaDetect}. The near-field detection of evanescent waves is the basis of how an STM functions. In an STM, a pointer is brought close to the sample and some of the evanescent waves hitting it are converted into propagating waves, which are then able to reach the detector.

Suppose we wish to resolve the sample on a scale $\sigma \ll \omega^{-1}$. This requires detecting evanescent modes with $k_x \sim \sigma^{-1}$, and corresponding $k_z = \sqrt{\omega^2 -k_x^2} \approx - i k_x$. Since the evanescent modes decay exponentially (\ref{eq:Evan}), the pointer tip must be placed no further than a distance $\Delta z \sim \sigma$ from the sample. Thus, STM allows resolution of features on arbitrarily short scales $\sigma$, {\sl regardless} of $\omega$. However, the new limitation is set by the distance of the pointer tip from the sample. To resolve on a scale $\sigma$ requires placing the tip no more than a distance $\sigma$ away from the sample so as to capture the evanescent modes.

\bibliographystyle{utcaps}
\bibliography{all}

%\end{comment}
\end{document}